\documentclass[aps,letterpaper,twocolumn,prl,footinbib,floatfix]{revtex4}
\usepackage{amsmath,amsfonts,amssymb}
\usepackage[scanall]{psfrag}
\usepackage{psfrag}
\usepackage{graphicx}
\usepackage{color}
\usepackage{tabularx}

\begin{document}

\newcommand{\of}[1]{\left( #1 \right)}
\newcommand{\sqof}[1]{\left[ #1 \right]}
\newcommand{\abs}[1]{\left| #1 \right|}
\newcommand{\avg}[1]{\left< #1 \right>}
\newcommand{\cuof}[1]{\left \{ #1 \right \} }
\newcommand{\bra}[1]{\left < #1 \right | }
\newcommand{\ket}[1]{\left | #1 \right > }
\newcommand{\pil}{\frac{\pi}{L}}
\newcommand{\bx}{\mathbf{x}}
\newcommand{\by}{\mathbf{y}}
\newcommand{\bk}{\mathbf{k}}
\newcommand{\bp}{\mathbf{p}}
\newcommand{\bl}{\mathbf{l}}
\newcommand{\bq}{\mathbf{q}}
\newcommand{\bs}{\mathbf{s}}
\newcommand{\psibar}{\overline{\psi}}
\newcommand{\svec}{\overrightarrow{\sigma}}
\newcommand{\dvec}{\overrightarrow{\partial}}
\newcommand{\bA}{\mathbf{A}}
\newcommand{\bdelta}{\mathbf{\delta}}
\newcommand{\bK}{\mathbf{K}}
\newcommand{\bQ}{\mathbf{Q}}
\newcommand{\bG}{\mathbf{G}}
\newcommand{\bw}{\mathbf{w}}
\newcommand{\bL}{\mathbf{L}}
\newcommand{\ohat}{\widehat{O}}
\newcommand{\up}{\uparrow}
\newcommand{\down}{\downarrow}
\newcommand{\MM}{\mathcal{M}}
\newcommand{\MN}{\mathcal{N}}
\newcommand{\MR}{\mathcal{R}}
\newcommand{\tW}{\tilde{W}}
\newcommand{\tX}{\tilde{X}}
\newcommand{\tY}{\tilde{Y}}
\newcommand{\tZ}{\tilde{Z}}
\newcommand{\tOm}{\tilde{\Omega}}
\newcommand{\barA}{\bar{\alpha}}

\author{George Grattan$^{1,2}$, Brandon A. Barton$^{1,3}$, Sean Feeney$^{1}$, Gianni Mossi$^{5,6}$, Pratik Patnaik$^{1,4}$, Jacob C. Sagal$^{1}$, Lincoln D. Carr$^{1,3,4}$, Vadim Oganesyan$^{7,8,9}$, and Eliot Kapit$^{1,4 *}$}
\address{$^1$ Quantum Engineering Program, Colorado School of Mines, 1523 Illinois St, Golden CO 80401}
\address{$^2$ Department of Computer Science, Colorado School of Mines, 1500 Illinois St, Golden CO 80401}
\address{$^3$ Department of Applied Mathematics and Statistics, Colorado School of Mines, 1500 Illinois St, Golden CO 80401}
\address{$^4$ Department of Physics, Colorado School of Mines, 1523 Illinois St, Golden CO 80401}

\address{$^5$ KBR, Inc., 601 Jefferson St., Houston, TX 77002, USA.}
\address{$^6$ Quantum Artificial Intelligence Lab. (QuAIL), NASA Ames Research Center, Moffett Field, CA 94035, USA.}
\address{$^7$ Department of Physics and Astronomy, College of Staten Island, CUNY, Staten Island, NY 10314, USA}
\address{$^8$ Physics program and Initiative for the Theoretical Sciences, The Graduate Center, CUNY, New York, NY 10016, USA}
\address{$^9$ Center for Computational Quantum Physics, Flatiron Institute, 162 5th Avenue, New York, NY 10010, USA}

\email{ekapit@mines.edu}

\title{Exponential acceleration of macroscopic quantum tunneling in a Floquet Ising model}

\begin{abstract}

The exponential suppression of macroscopic quantum tunneling (MQT) in the number of elements to be reconfigured is an essential element of broken symmetry phases. Slow MQT is also a core bottleneck in quantum algorithms, such as traversing an energy landscape in optimization, and adiabatic state preparation more generally. In this work, we demonstrate the possibility to accelerate MQT through Floquet engineering with the application of a uniform, high frequency transverse drive field. Using the ferromagnetic phase of the transverse field Ising model in one and two dimensions as a prototypical example, we identify three qualitatively distinct regimes as a function of drive strength: (i) for weak drive, the system exhibits exponentially slow MQT alongside robust magnetic order, as expected; (ii) at intermediate drive strength, we find polynomial decay of rates alongside vanishing magnetic order consistent with critical or paramagnetic state; (iii) at very strong drive strengths both the tunnelling rate and time-averaged magnetic order remain finite 
with increasing system size. We support these claims with extensive full wavefunction and matrix-product state numerical simulations, and theoretical analysis.  An experimental test of these results  presents a technologically important and novel scientific question accessible on NISQ-era quantum computers.
\end{abstract}

\maketitle

The idea of single-particle quantum tunneling through a barrier is well-known since the 1920s~\cite{gurney1929quantum,hund1927interpretation} and is found in practical technologies such as the tunneling diode~\cite{esaki1958new}.  Common examples of macroscopic quantum tunneling (MQT) have typically built on this concept~\cite{zhao2017macroscopic}, including the Josephson junction, a building block of quantum information systems, and can be modeled by e.g. the Lipkin-Meshkov-Glick model~\cite{lipkin1965validity}. 
A long-time goal of such models is to move beyond the Josephson regime, in which $N$ bosons or Cooper pairs move fluidly between two dominant single-particle states a particle at a time, into the Fock regime, in which all particles can collectively tunnel from one extreme to the other -- the $|N,0\rangle$ and $|0,N\rangle$ NOON state. 
This kind of tunneling is hard to observe because the tunneling time is exponentially long in the number of particles. One way to understand this exponentially long time is to calculate the energy splitting between symmetric and anti-symmetric states $|N,0\rangle \pm |0,N\rangle$, which is exponentially small in the Fock regime. The tunneling time can be estimated as $\hbar$ over this energy splitting. Such concepts are the basis of symmetry breaking and are famously cited in Anderson's paper, ``More is Different''~\cite{anderson1972more}, where he uses the example of left and right handed sugar as the two extremes.  In this case, the tunneling time is longer than the lifetime of the universe.  However, the current quantum computing paradigm and NISQ device availability offer a new opportunity to rexamine such foundational questions, due to their high level of many-body control in both time and space.  Utilizing many-body control in the form of \emph{symphonic tunneling} \cite{mossi2023embedding} on the transverse field Ising model (TFIM), in this Letter we establish a complexity transition in MQT from exponential suppression to a polynomial scaling in the number of particles.

\begin{figure*}\includegraphics[width=0.24\textwidth]{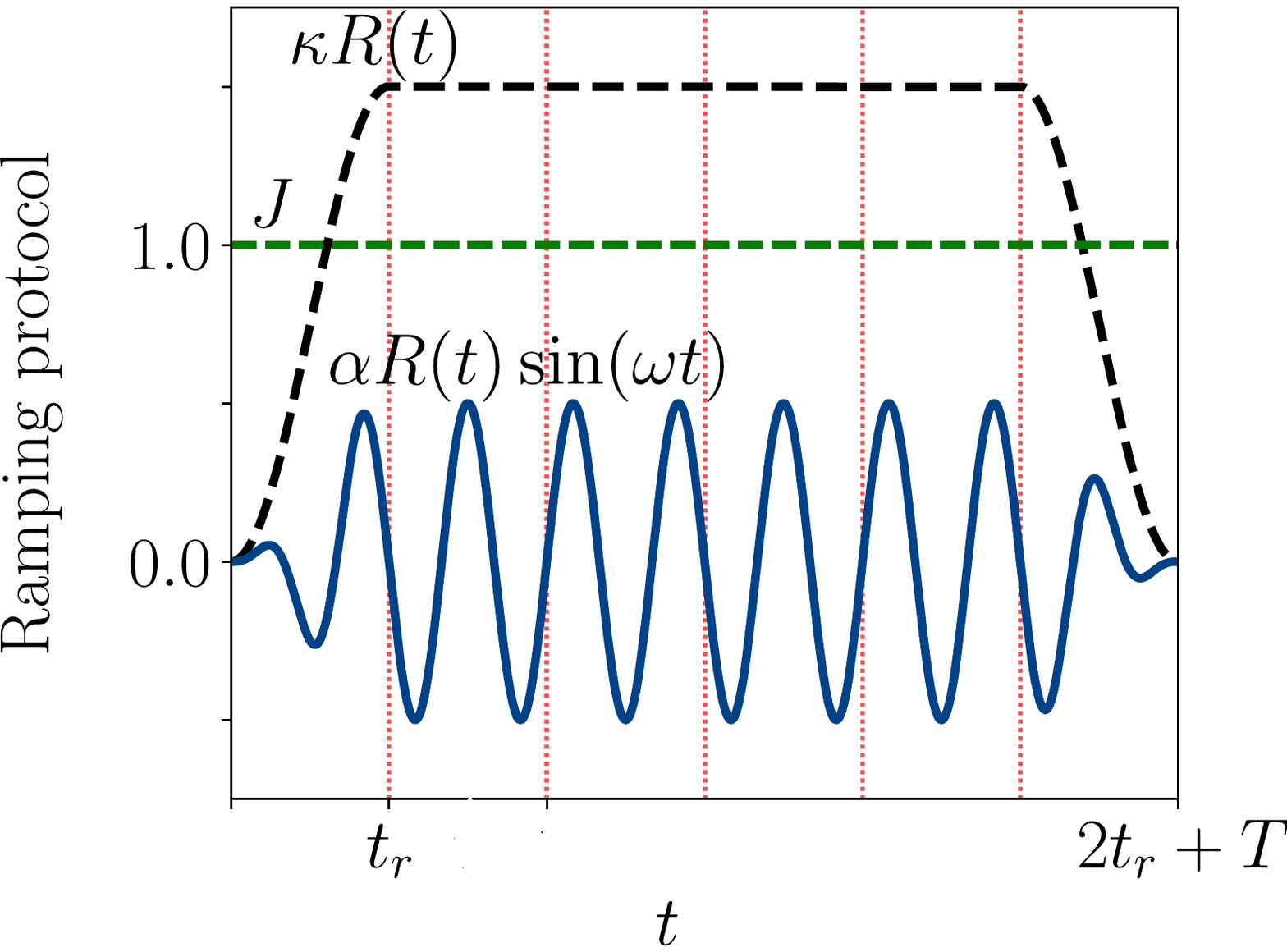}\hfill
    \includegraphics[width=0.48\textwidth]{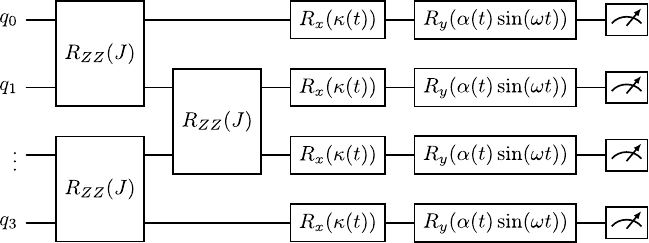}\hfill
    \includegraphics[width=0.24\textwidth]{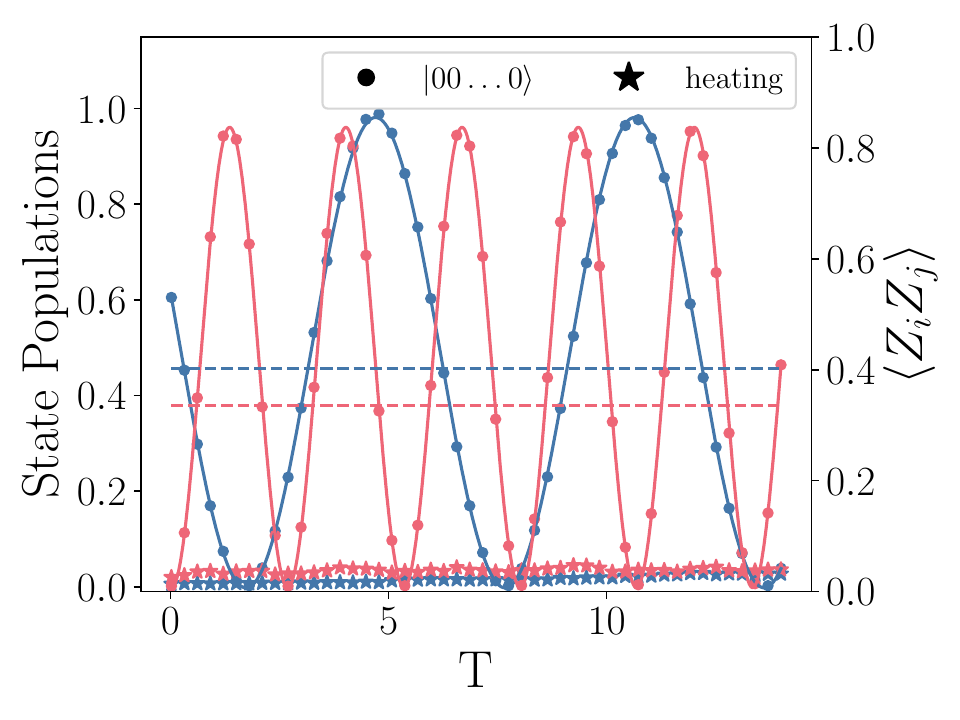}
    \caption{ (left) Adiabatic ramping protocols in for the coefficients $\{J,\kappa, \alpha\}$ in Eq. (\ref{eq:2d_ising}). Both the $X$ (DC) and $Y$ (AC) fields are smoothly ramped up $R(t<t_r)=\sin^2 0.5 \pi t/t_r$ and similarly down from $(t_r + T) \rightarrow (2t_r + T)$ with the same sinusoidal profile, $R(t)$. (center) The circuit is repeated with varying plateau times $T$ to measure the period of Rabi oscillations, $2\pi/\Omega_R\ll T$. (right) Time traces used to extract the oscillation period $\Omega_R$, for a $4 \times 7$ 2d  geometry, with $\kappa = 2.5$ and two choices of $\alpha/f$. We plot the final population of $\ket{00 \ldots 0}$ (after ramping fields down), for a series of plateau times $T$. The average ferromagnetic correlator $\avg{Z_i Z_j}$ (dashed lines) is measured repeatedly during evolution for drive strengths $\alpha / f = 0.1$ (blue) and 0.14 (red) with sinusoidal fits (solid curves) and an estimate of heating $1-\abs{\left < \psi_f | 00 \ldots 0 \right>}^2-\abs{\left < \psi_f | 11 \ldots 1 \right>}^2$ (stars near zero).}
    \label{fig:TFIM_ramp}
\end{figure*}

Prior to the wide availability of NISQ quantum computing devices and the spread of the quantum circuit paradigm, many researchers looked for ways to observe MQT, such as scattering solitons or ``lumps'' of bosons on a barrier~\cite{weiss2012elastic}, physically manipulating complicated traps in quantum simulator experiments~\cite{shin2004distillation,potnis2017interaction}, hybridizing modes to access new avenues for quantum control~\cite{cristofolini2012coupling,carusotto2013quantum}, or driving non- or weakly-interacting systems~\cite{grifoni1998driven,eckardt2017colloquium}.  A particularly famous example is the beam splitter experiment of Markus Arndt~\cite{arndt1999wave}, in which fullerenes take two simultaneous paths.  Although these experiments are in a certain sense macroscopic~\cite{nimmrichter2013macroscopicity}, ultimately they are mean-field like superpositions of center of mass degrees of freedom~\cite{zhao2017macroscopic}, just like Josephson tunneling.  

However, no one to date has determined a realistic way to achieve MQT of large numbers of strongly-interacting quantum elements in a Fock-like regime, and even controlling small numbers of weakly interacting fermions has been a grand challenge only recently accomplished in microtraps~\cite{serwane2011deterministic}.  This is a desirable goal in order to traverse energy landscapes, for instance, to find a ground state of a spin glass \cite{farhi2014quantum,albash2018adiabatic}. MQT physics forms a key source of potential quantum advantage, and bottleneck process for quantum optimization \cite{altshulerkrovi2010,knysh2016}.  
While we have previously explored multi-tone drives to accelerate MQT scaling\cite{tang2021unconventional,mossi2023embedding,kapit2021noise}, in this work we focus on a single frequency \emph{Floquet engineering} approach \cite{oka2019floquet}, which we implement via Trotterization\cite{trotter1959product} in discrete space-time, i.e. on a quantum circuit model in one and two dimensional lattices of qubits. In addition to the expected ferromagnetic and crossover (critical or paramagnetic) behaviors observed at weak and intermediate drives, we discover an unusual revival of ferromagnetic order at very strong drives accompanied by comparatively fast MQT. 
The quantum Ising model we study is enriched with a time dependent y-field of strength $\alpha$ and angular frequency $\omega$
\begin{eqnarray}\label{eq:2d_ising}
    H = -J \sum_{\langle ij \rangle } Z_i Z_j - \kappa \sum_{i}  X_i + \alpha \sin{{\omega t}} \sum_{i} Y_i,
\end{eqnarray}
where familiar static couplings $J, \kappa>0$  denote ferromagnetic spin-spin interaction and transverse field, respectively. 
The phase transition to a paramagnetic state occurs (for $\alpha=0$) at $\kappa_c=\kappa/J = 1$ ($\simeq 3$) in 1D (2D)~\cite{schmitt2022quantum}. For $\kappa<\kappa_c$, the ground state manifold is a doublet of symmetric and antisymmetric superpositions of macroscopic polarization states with energy splitting $\Omega_0 \of{N}$, often referred to as a Rabi frequency given the simplicity of the two-level dynamics that ensues.
This splitting $\Omega_0 \of{N}$--the inverse of which sets the MQT time to mix the two ferromagnetic states--decays exponentially in $N$. In particular, in 1D we obtain analytically (see Sec. B in Supp. Materials (SM))
\begin{eqnarray}
\Omega_0 \of{N} \propto \frac{\kappa}{\sqrt{N}} \of{\frac{\kappa}{J}}^{N-1}.
\label{eq:1DFA}
\end{eqnarray}
We are not aware of similar calculation in 2D but the result should scale exponentially in $N$, i.e. area-like. 

To measure the tunneling rate we initialize the system (Fig.~\ref{fig:TFIM_ramp}, left) in one of the FM groundstates of the TFIM for $\kappa=\alpha=0$, and smoothly ramp up off-diagonal terms $\kappa$ and $\alpha$.  This creates a coherent magnetization reversal (Rabi) oscillation which we measure by varying the duration of the plateau and fitting the probability of magnetization reversal after ramping down to a simple cosine profile (Fig.~\ref{fig:TFIM_ramp}, right).
Magnetic order is inferred from the time average of the two-point correlation $\avg{Z_i Z_j}$ over the entire plateau region evaluated at large spatial separations. Our Trotterized simulations closely model how a gate-based quantum computer would approximate the continuum time evolution of the problem (Fig.~\ref{fig:TFIM_ramp}, center), i.e. with a sufficiently small time step (see SM).

\begin{figure*}
\includegraphics[width=0.32\textwidth]{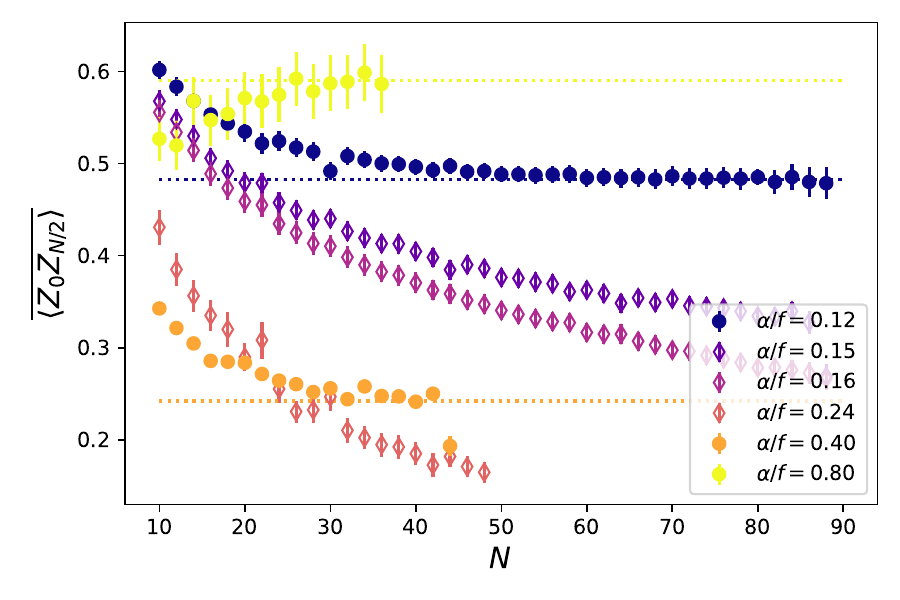}
\includegraphics[width=0.32\textwidth]{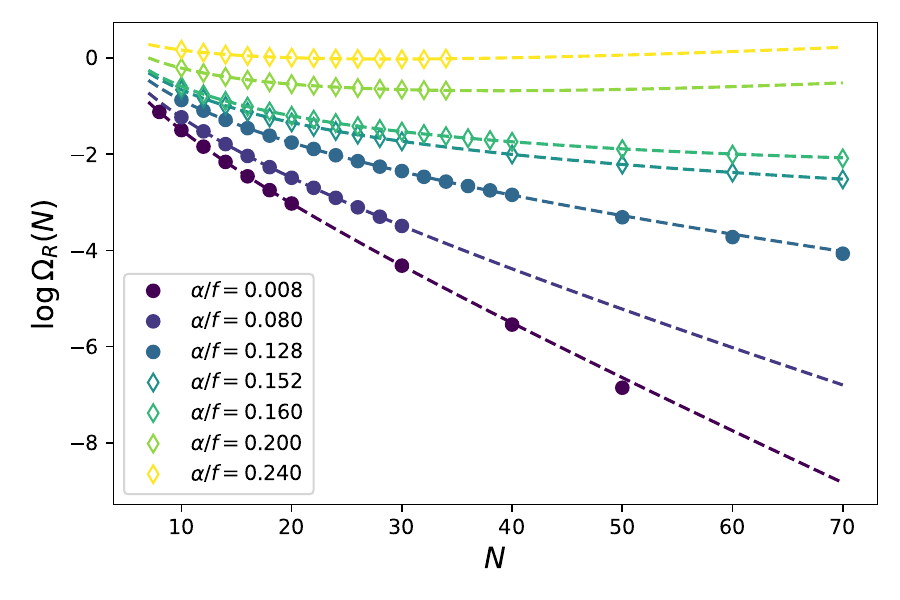}
\includegraphics[width=0.32\textwidth]{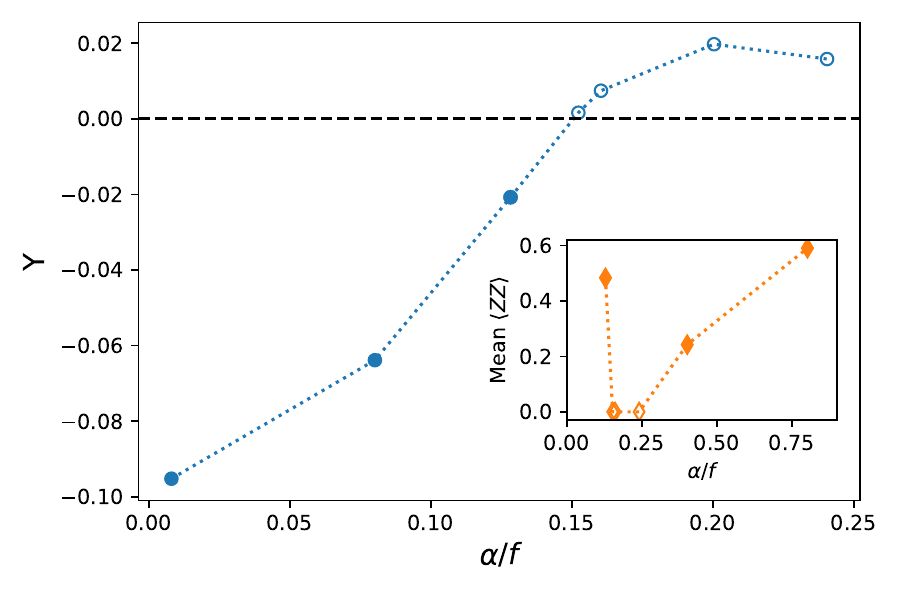}
 \caption{Evolution of order and MQT in the driven 1D quantum Ising model from weak ($\alpha/f<0.15$) to strong ($\alpha/f\geq 0.4$) driving, simulated with time-evolved block decimation (TEBD), with transverse field $\kappa = 0.9$. Left: antipodal magnetic order parameter -- note the saturation of order at large separation for weak and strong driving regime, with apparent decay in-between. Drive parameters for which simulations predict magnetic order is  asymptotically zero (finite) are plotted with hollow diamonds (solid circles). Center: exponential decay of the Rabi freq. associated with MQT for weak drives gives way to non-exponential, nearly flat behavior vs. $N$ as $\alpha/f$ increases. Drive parameters for which the best fit is exponential decay are plotted with solid circles; hollow diamonds indicate polynomial decay or constant $\Omega_R$. Right: Decay ("difficulty") exponents $\Upsilon$ for $\Omega_R(N)$. 
     Solid dots indicate data points where $\Omega_R$ decays exponentially, while empty circles are compatible with subexponential decay. Inset: time-averaged values for the two-point correlator from Fig. 2.
     Filled diamonds indicate values of $\alpha/f$ where we observe saturation of $\langle Z_0Z_{L/2} \rangle$ in the range of system sizes observed. Empty diamonds label cases where no such saturation is observed, which we conservatively interpret a zero magnetization.
 }
    \label{fig:1d}
\end{figure*}

Strong AC drive in Eq. \ref{eq:2d_ising} results in novel terms in the effective Floquet Hamiltonian (see Sec. D in SM) -- steering coherent correlated many-body behavior using such dynamically generated Hamiltonians is commonly referred to as \emph{Floquet engineering} (FE)\cite{bukov2016schrieffer,rudner2020floquet}. 
The apparent dramatic renormalization of MQT 
has not previously been realized through FE, to the best of our knowledge. Since heating is the common concern in FE we took special care to mitigate it by 
implementing a scaling limit in which the drive frequency and amplitude both increase logarithmically with $N$. This choice is motivated by the expectation that heating rates from high-frequency drives decay exponentially in $\omega$ \cite{abanin2015exponentially,mallayya2019heating,else2020long,machado2020long,zhao2021random,shkedrov2022absence,ikeda2023robust}, but increase only quadratically with drive amplitude \cite{mallayya2019heating}. We thus redefine the Floquet controls as $\alpha$, $f$ with
\begin{equation}
\alpha \equiv \alpha_s \log N,\:\omega =2\pi f \equiv 2\pi f_s \log N
\end{equation} 
such that the state evolves as
\begin{equation}
    \ket{\psi(t+dt)} = \exp \of{-2 \pi i dt H(t)} \ket{\psi \of{t}}.
\end{equation} The  leading corrections to the Floquet Hamiltonian are generated as a power series in $\alpha/f=\alpha_s/f_s$, and therefore remain constant in $N$ (also, note the convention of extra $2\pi$ - see SM, Sec. D). This log-over-log limit hits an empirical sweet spot where the driving frequency increases very gradually, but the system can evolve for polynomially  long times (or even longer) before meaningfully heating.  In practical terms, such logarithmic scaling limit translates into logarithmic increase in circuit depth on a digital quantum computer and should be readily reacheable. Analog implementations will likely be more nuanced and mindful of specific relaxation mechanisms (see SM, Sec. D for added discussion).
Smooth ramp profiles (see Fig. \ref{fig:TFIM_ramp}) ensure adiabatic loading of the many-body state (see SM Sec. E for quantitative study of Floquet adiabaticity).  
We note finally that we cannot measure heating during evolution as we do not compute the eigenstates of the AC-dressed system (except in SM, Sec. E), and thus can only measure it after the transverse terms are ramped down. Since the system is closed, a final measurement is sufficient to detect if the system left the ground state manifold (Fig.~\ref{fig:TFIM_ramp}, right). Throughout this work heating was monitored vigilantly and suppressed through a combination of measures described in this paragraph. 

\begin{figure}[b]
\includegraphics[width=\columnwidth]{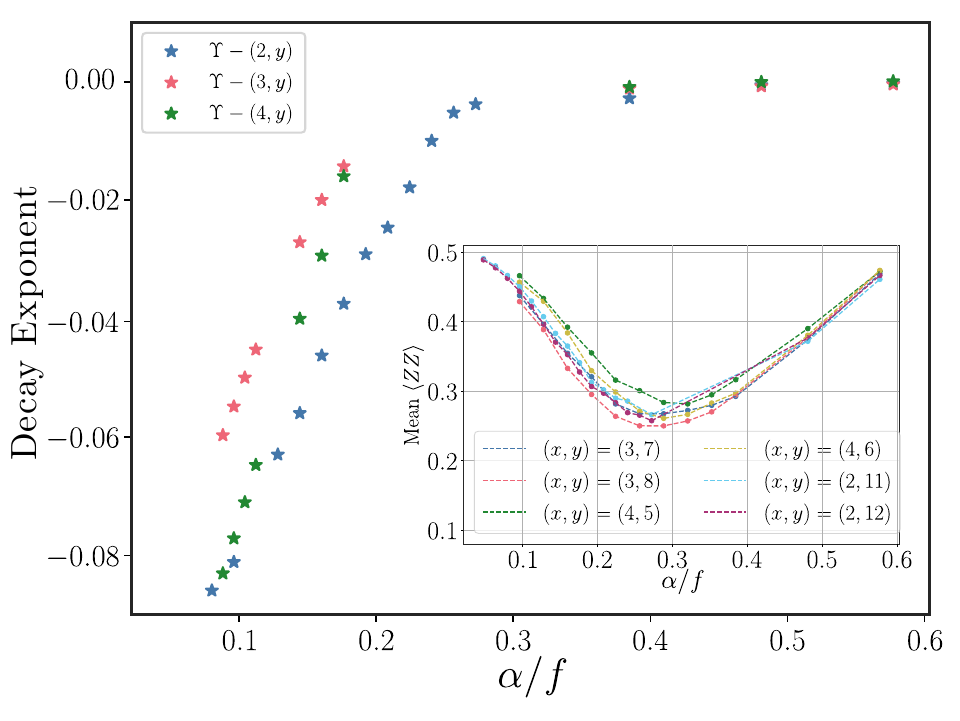}
 \caption{Evolution of order and MQT in the 2D driven quantum Ising model, for various geometries with $12 \leq N \leq 28$, with the time-averaged magnetic order parameter plotted in the inset for four relatively large system sizes. 
 As the decay exponent $\Upsilon \to 0$, the polynomial prefactor in $\Omega_R$ reduces to nearly constant scaling in the very strong drive limit. We used a transverse field $\kappa = 2.5$ for $3 \times y$ and $4 \times y$ geometries, and 1.75 for $2 \times y$.
 }
\label{fig:2d}
\end{figure}

The key results of our simulations are shown in Fig.~\ref{fig:1d} (1D) and Fig. \ref{fig:2d}(2D). They document the evolution of long-range magnetic order and also of the Rabi precession frequency of macroscopic magnetization. Most remarkably, 1D and 2D appear phenomenologically identical. We can identify three distinct regimes:

\noindent
(i) The \emph{weak driving} regime ($\alpha/f\lesssim 0.2$ in 2D and $\alpha/f< 0.15$ in 1D): here $\Omega_R \propto N^{-c} \exp \of{-\Upsilon N}$. The scaling exponent $\Upsilon$ is reduced by the AC drive but the splitting still scales exponentially. At long distances $\avg{Z_i Z_j}$ approaches  a non-zero constant, see Figs.~\ref{fig:1d} and \ref{fig:2d}. This is the same qualitative behavior as the DC (undriven) problem, and also provides a test case for quality control of our two main simulation methods. 

\noindent
(ii) The \emph{crossover} regime ($0.2\lesssim\alpha/f\lesssim 0.3$)
is characterized by monotonic decay of spin correlations consistent with absence of magnetic order;

\noindent
(iii) The \emph{strong driving} regime ($\alpha/f \gtrsim 0.3$ in 2D, $\alpha/f \geq 0.4$ in 1D): $\avg{Z_i Z_j}$ begins to increase and becomes constant as a function of distance. $\Omega_R$ likewise becomes approximately constant with system size, though the total time required to tunnel between states is polynomial because a polynomially long ramp time is required to turn on the transverse terms while avoiding heating. The physics of this regime may be traced (see SM Sec. E) to so-called Floquet "micromotion", i.e. fast intra-period dynamics, which also leads to strong dependence of observables on the precise protocol used for time averaging, e.g. we noticed larger error bars in Fig.\ref{fig:2d} (left). The strong driving regime here is particularly surprising. As $\alpha$ increases, we observe this behavior regime begins close to $\alpha/f \sim 0.3\sim 0.4$ 
and, importantly, for a variety of $\kappa$ values (not shown), in contrast to regimes (i) and (ii) which are much more sensitive to $\kappa/J$. Our approximate analytic calculation is able to capture the variation of Rabi frequency in the weak drive regime (i) and also the onset of (ii) in a modified version of Eq.~(\ref{eq:1DFA}) (see SM, Sec. C), This derivation does not capture the restoration of order at strong driving, even with accurately extracted (numerically) Floquet Hamiltonian (see SM, Sec. D), thus presenting an open problem which likely requires a more nuanced treatment of Floquet micromotion (see also SM, Sec. E) and possibly other effects.

\begin{figure}
\includegraphics[width=\columnwidth]{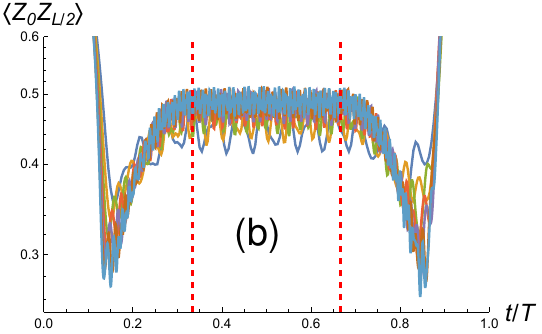}
\caption{Time-averaged magnetic order as probed with antipodal correlator in 1D chains of length 8 to 20 (dark blue, gold, green, red, purple, brown, light blue) at strong AC drive. Red dashed lines demarcate waiting time in-between the on-/off-ramp whereby both DC and AC transverse fields are simultaneously turned on. Contrary to conventional behavior the order parameter achieves its minimal value \emph{during} the ramp and then recovers rather dramatically on the plateaus.
}
\label{fig:revival}
\end{figure}

We can further explore the non-equilibrium nature of the strong driving regime by examining the temporal evolution of the same antipodal correlator we used to define long range order in 1D. The existence of paramagnetic intermediate $\alpha$ regime is probed during the ramp, as the magnetic order appears to nearly collapse but then revive briskly in time for the plateau, see Fig. \ref{fig:revival}.  
As mentioned above, the Rabi oscillation rate $\Omega_R$ measured from varying the plateau time is approximately constant with system size. $\Omega_R$ in this limit is a continuous function of $\kappa$, $\alpha$ and $f$ and not a simple multiple of any of them. This is in stark contrast to simply using a large DC transverse field (e.g. $\kappa > \kappa_c$), where no such order restoration is seen and magnetic order decays exponentially with $N$ in the plateau region. However, the ramping time to reach the plateau without heating the system does increase with $N$, empirically as $O \of{N^2}$ in 1D, which combined with the decreasing minimum ferromagnetic order parameter (during the ramp) suggests to us that we cross at least one, possibly two, phase transitions en route to the ferromagnetic state at strong drive. Taken together with the vastly more complicated temporal dynamics of this correlator on the plateau, which shows strong high frequency components averaged over and not shown in the Fig. \ref{fig:revival}, and  the inability of the quasi-equilibrium average Hamiltonian approach (see SM) to capture the restoration of order parameter, our numerical results suggest the importance of non-equilibrium effects, such as prethermalization.

A natural and obvious extension of the results of this Letter is accelerated quantum optimization, though we caution that the spin glass case is much more complex (for example, no longer being a simple ferromagnet, the pair-flip terms generated will no longer be sign-definite and can interfere destructively with the DC transverse field). Applying a simple uniform AC field everywhere is not expected to produce significant benefits for such problems, necessitating a more sophisticated approach. That said, it may be helpful in mitigating some of the slowdowns associated with minor embedding \cite{mossi2023embedding} in analog quantum optimization, by inducing fast MQT in the chains that embed logical qubits in 2D systems. 

Another natural next step would be to test these predictions on real quantum hardware. The feasibility of such an experiment is bolstered by the recent Quantum Utility experiment of the IBM team \cite{kim2023evidence}, which demonstrated accurate simulation of time evolution in a TFIM with up to 127 qubits, by combining multiple error mitigation techniques.  Our discoveries here present a concrete, physically relevant, test: Floquet engineering of fast MQT coexisting with broken symmetry. The apparent universality of our results strongly suggests that this phenomenon is real at large scales in higher dimensions, but one of course needs to do an experiment on quantum computing hardware to be certain. The oscillation periods in the strong drive regimes in Figs.~\ref{fig:1d} and \ref{fig:2d} are empirically quite short, in the range of 10-20 Trotter steps.  We estimate that with a bit more effort put into fine tuning the ramping profile,  system parameters, and circuit, perhaps using novel methods such as \cite{eckstein2023large}, present or near-term quantum hardware could simultaneously extract the scaling of the oscillation period and ferromagnetic order parameter for a hundred or more qubits. This would be one of the first uses of a quantum computer to answer a question of genuine scientific interest at beyond-classical scales, a significant milestone in the progress of quantum information science.

In conclusion, using a mix of theoretical arguments and large scale numerical simulations, with the transverse field Ising ferromagnet as a model system, we have shown that strong, high frequency AC drives can dramatically increase macroscopic quantum tunneling rates, inducing a crossover from exponential to polynomial or even constant scaling with system size. We further observed an unusual but consistent \emph{increase} in the time averaged magnetic order with very strong driving, coexisting with fast MQT. This new dynamical phase is not explained by our analytical theory and deserves further analysis. As large MQT events form a key bottleneck in quantum algorithms, novel methods to accelerate them based on extensions of this work could have broad impact.

\begin{acknowledgments}
We would like to thank Marin Bukov, David Huse, Tom Iadecola, Peter Orth, Anatoli Polkovnikov for useful discussions. We would also like to thank Takuto Komatsuki and Joey Liu for support with HPC calculations. This work was supported in part by the DARPA Reversible Quantum Machine Learning and Simulation program under contract HR00112190068, as well as by National Science Foundation grants PHY-1653820, PHY-2210566, DGE-2125899. G.G. was additionally supported by ARO grant No. W911NF-18-1-0125. E.K's advisory role in the project, and G.G's quantum simulations, were supported by the U.S. Department of Energy, Office of Science, National Quantum Information Science Research Centers, Superconducting Quantum Materials and Systems Center (SQMS) under contract number DE-AC02-07CH11359, with G.M.'s work funded under the NASA-DOE interagency agreement SAA2-403602 governing NASA's work as part of the SQMS center. G.M. is a KBR employee working under the Prime Contract No. 80ARC020D0010
with the NASA Ames Research Center, and additionally acknowledges support from DARPA under IAA 8839, Annex 128. Resources supporting this work were also provided by the NASA High-End Computing (HEC) Program through the NASA Advanced Supercomputing (NAS) Division at the Ames Research Center. Many of the numerical simulations in this work were performed with a generous grant of HPC access from the Fujitsu Corporation. Part of this research was performed while the one of the authors was visiting the Institute for Pure and Applied Mathematics (IPAM), which is supported by the National Science Foundation (Grant No. DMS-1925919). The Flatiron Institute is a division of the Simons Foundation.
\end{acknowledgments}

\bibliography{fullbib}

\begin{thebibliography}{9}
\expandafter\ifx\csname natexlab\endcsname\relax\def\natexlab#1{#1}\fi
\expandafter\ifx\csname bibnamefont\endcsname\relax
  \def\bibnamefont#1{#1}\fi
\expandafter\ifx\csname bibfnamefont\endcsname\relax
  \def\bibfnamefont#1{#1}\fi
\expandafter\ifx\csname citenamefont\endcsname\relax
  \def\citenamefont#1{#1}\fi
\expandafter\ifx\csname url\endcsname\relax
  \def\url#1{\texttt{#1}}\fi
\expandafter\ifx\csname urlprefix\endcsname\relax\def\urlprefix{URL }\fi
\providecommand{\bibinfo}[2]{#2}
\providecommand{\eprint}[2][]{\url{#2}}

\bibitem[{\citenamefont{Bravyi et~al.}(2011)\citenamefont{Bravyi, DiVincenzo,
  and Loss}}]{bravyi2011schrieffer}
\bibinfo{author}{\bibfnamefont{S.}~\bibnamefont{Bravyi}},
  \bibinfo{author}{\bibfnamefont{D.~P.} \bibnamefont{DiVincenzo}},
  \bibnamefont{and} \bibinfo{author}{\bibfnamefont{D.}~\bibnamefont{Loss}},
  \bibinfo{journal}{Annals of physics} \textbf{\bibinfo{volume}{326}},
  \bibinfo{pages}{2793} (\bibinfo{year}{2011}).

\bibitem[{\citenamefont{Pietracaprina et~al.}(2016)\citenamefont{Pietracaprina,
  Ros, and Scardicchio}}]{pietracaprina2016forward}
\bibinfo{author}{\bibfnamefont{F.}~\bibnamefont{Pietracaprina}},
  \bibinfo{author}{\bibfnamefont{V.}~\bibnamefont{Ros}}, \bibnamefont{and}
  \bibinfo{author}{\bibfnamefont{A.}~\bibnamefont{Scardicchio}},
  \bibinfo{journal}{Physical Review B} \textbf{\bibinfo{volume}{93}},
  \bibinfo{pages}{054201} (\bibinfo{year}{2016}).

\bibitem[{\citenamefont{Baldwin et~al.}(2016)\citenamefont{Baldwin, Laumann,
  Pal, and Scardicchio}}]{baldwinlaumann2016}
\bibinfo{author}{\bibfnamefont{C.}~\bibnamefont{Baldwin}},
  \bibinfo{author}{\bibfnamefont{C.}~\bibnamefont{Laumann}},
  \bibinfo{author}{\bibfnamefont{A.}~\bibnamefont{Pal}}, \bibnamefont{and}
  \bibinfo{author}{\bibfnamefont{A.}~\bibnamefont{Scardicchio}},
  \bibinfo{journal}{Physical Review B} \textbf{\bibinfo{volume}{93}},
  \bibinfo{pages}{024202} (\bibinfo{year}{2016}).

\bibitem[{\citenamefont{Baldwin et~al.}(2017)\citenamefont{Baldwin, Laumann,
  Pal, and Scardicchio}}]{baldwinlaumann2017}
\bibinfo{author}{\bibfnamefont{C.}~\bibnamefont{Baldwin}},
  \bibinfo{author}{\bibfnamefont{C.}~\bibnamefont{Laumann}},
  \bibinfo{author}{\bibfnamefont{A.}~\bibnamefont{Pal}}, \bibnamefont{and}
  \bibinfo{author}{\bibfnamefont{A.}~\bibnamefont{Scardicchio}},
  \bibinfo{journal}{Physical Review Letters} \textbf{\bibinfo{volume}{118}},
  \bibinfo{pages}{127201} (\bibinfo{year}{2017}).

\bibitem[{\citenamefont{Scardicchio and
  Thiery}(2017)}]{scardicchio2017perturbation}
\bibinfo{author}{\bibfnamefont{A.}~\bibnamefont{Scardicchio}} \bibnamefont{and}
  \bibinfo{author}{\bibfnamefont{T.}~\bibnamefont{Thiery}},
  \bibinfo{journal}{arXiv preprint arXiv:1710.01234}  (\bibinfo{year}{2017}).

\bibitem[{\citenamefont{Baldwin and Laumann}(2018)}]{baldwin2018quantum}
\bibinfo{author}{\bibfnamefont{C.}~\bibnamefont{Baldwin}} \bibnamefont{and}
  \bibinfo{author}{\bibfnamefont{C.}~\bibnamefont{Laumann}},
  \bibinfo{journal}{Physical Review B} \textbf{\bibinfo{volume}{97}},
  \bibinfo{pages}{224201} (\bibinfo{year}{2018}).

\bibitem[{\citenamefont{Kapit and Oganesyan}(2022)}]{kapit2022small}
\bibinfo{author}{\bibfnamefont{E.}~\bibnamefont{Kapit}} \bibnamefont{and}
  \bibinfo{author}{\bibfnamefont{V.}~\bibnamefont{Oganesyan}},
  \bibinfo{journal}{arXiv preprint arXiv:2212.04588}  (\bibinfo{year}{2022}).

\bibitem[{\citenamefont{Morningstar et~al.}(2023)\citenamefont{Morningstar,
  Huse, and Khemani}}]{morningstar2023universality}
\bibinfo{author}{\bibfnamefont{A.}~\bibnamefont{Morningstar}},
  \bibinfo{author}{\bibfnamefont{D.~A.} \bibnamefont{Huse}}, \bibnamefont{and}
  \bibinfo{author}{\bibfnamefont{V.}~\bibnamefont{Khemani}},
  \bibinfo{journal}{Phys. Rev. B} \textbf{\bibinfo{volume}{108}},
  \bibinfo{pages}{174303} (\bibinfo{year}{2023}),
  \urlprefix\url{https://link.aps.org/doi/10.1103/PhysRevB.108.174303}.

\bibitem[{\citenamefont{Schindler and
  Bukov}(2023)}]{schindler2023counterdiabatic}
\bibinfo{author}{\bibfnamefont{P.~M.} \bibnamefont{Schindler}}
  \bibnamefont{and} \bibinfo{author}{\bibfnamefont{M.}~\bibnamefont{Bukov}},
  \bibinfo{journal}{arXiv preprint arXiv:2310.02728}  (\bibinfo{year}{2023}).

\end{thebibliography}


\begin{thebibliography}{37}
\expandafter\ifx\csname natexlab\endcsname\relax\def\natexlab#1{#1}\fi
\expandafter\ifx\csname bibnamefont\endcsname\relax
  \def\bibnamefont#1{#1}\fi
\expandafter\ifx\csname bibfnamefont\endcsname\relax
  \def\bibfnamefont#1{#1}\fi
\expandafter\ifx\csname citenamefont\endcsname\relax
  \def\citenamefont#1{#1}\fi
\expandafter\ifx\csname url\endcsname\relax
  \def\url#1{\texttt{#1}}\fi
\expandafter\ifx\csname urlprefix\endcsname\relax\def\urlprefix{URL }\fi
\providecommand{\bibinfo}[2]{#2}
\providecommand{\eprint}[2][]{\url{#2}}

\bibitem[{\citenamefont{Gurney and Condon}(1929)}]{gurney1929quantum}
\bibinfo{author}{\bibfnamefont{R.~W.} \bibnamefont{Gurney}} \bibnamefont{and}
  \bibinfo{author}{\bibfnamefont{E.~U.} \bibnamefont{Condon}},
  \bibinfo{journal}{Physical Review} \textbf{\bibinfo{volume}{33}},
  \bibinfo{pages}{127} (\bibinfo{year}{1929}).

\bibitem[{\citenamefont{Hund}(1927)}]{hund1927interpretation}
\bibinfo{author}{\bibfnamefont{F.}~\bibnamefont{Hund}}, \bibinfo{journal}{z.
  Phys} \textbf{\bibinfo{volume}{43}}, \bibinfo{pages}{805}
  (\bibinfo{year}{1927}).

\bibitem[{\citenamefont{Esaki}(1958)}]{esaki1958new}
\bibinfo{author}{\bibfnamefont{L.}~\bibnamefont{Esaki}},
  \bibinfo{journal}{Physical review} \textbf{\bibinfo{volume}{109}},
  \bibinfo{pages}{603} (\bibinfo{year}{1958}).

\bibitem[{\citenamefont{Zhao et~al.}(2017)\citenamefont{Zhao, Alcala, McLain,
  Maeda, Potnis, Ramos, Steinberg, and Carr}}]{zhao2017macroscopic}
\bibinfo{author}{\bibfnamefont{X.}~\bibnamefont{Zhao}},
  \bibinfo{author}{\bibfnamefont{D.~A.} \bibnamefont{Alcala}},
  \bibinfo{author}{\bibfnamefont{M.~A.} \bibnamefont{McLain}},
  \bibinfo{author}{\bibfnamefont{K.}~\bibnamefont{Maeda}},
  \bibinfo{author}{\bibfnamefont{S.}~\bibnamefont{Potnis}},
  \bibinfo{author}{\bibfnamefont{R.}~\bibnamefont{Ramos}},
  \bibinfo{author}{\bibfnamefont{A.~M.} \bibnamefont{Steinberg}},
  \bibnamefont{and} \bibinfo{author}{\bibfnamefont{L.~D.} \bibnamefont{Carr}},
  \bibinfo{journal}{Physical Review A} \textbf{\bibinfo{volume}{96}},
  \bibinfo{pages}{063601} (\bibinfo{year}{2017}).

\bibitem[{\citenamefont{Lipkin et~al.}(1965)\citenamefont{Lipkin, Meshkov, and
  Glick}}]{lipkin1965validity}
\bibinfo{author}{\bibfnamefont{H.~J.} \bibnamefont{Lipkin}},
  \bibinfo{author}{\bibfnamefont{N.}~\bibnamefont{Meshkov}}, \bibnamefont{and}
  \bibinfo{author}{\bibfnamefont{A.}~\bibnamefont{Glick}},
  \bibinfo{journal}{Nuclear Physics} \textbf{\bibinfo{volume}{62}},
  \bibinfo{pages}{188} (\bibinfo{year}{1965}).

\bibitem[{\citenamefont{Anderson}(1972)}]{anderson1972more}
\bibinfo{author}{\bibfnamefont{P.~W.} \bibnamefont{Anderson}},
  \bibinfo{journal}{Science} \textbf{\bibinfo{volume}{177}},
  \bibinfo{pages}{393} (\bibinfo{year}{1972}).

\bibitem[{\citenamefont{Mossi et~al.}(2023)\citenamefont{Mossi, Oganesyan, and
  Kapit}}]{mossi2023embedding}
\bibinfo{author}{\bibfnamefont{G.}~\bibnamefont{Mossi}},
  \bibinfo{author}{\bibfnamefont{V.}~\bibnamefont{Oganesyan}},
  \bibnamefont{and} \bibinfo{author}{\bibfnamefont{E.}~\bibnamefont{Kapit}},
  \bibinfo{journal}{arXiv preprint arXiv:2306.10632}  (\bibinfo{year}{2023}).

\bibitem[{\citenamefont{Weiss and Castin}(2012)}]{weiss2012elastic}
\bibinfo{author}{\bibfnamefont{C.}~\bibnamefont{Weiss}} \bibnamefont{and}
  \bibinfo{author}{\bibfnamefont{Y.}~\bibnamefont{Castin}},
  \bibinfo{journal}{Journal of Physics A: Mathematical and Theoretical}
  \textbf{\bibinfo{volume}{45}}, \bibinfo{pages}{455306}
  (\bibinfo{year}{2012}).

\bibitem[{\citenamefont{Shin et~al.}(2004)\citenamefont{Shin, Saba, Schirotzek,
  Pasquini, Leanhardt, Pritchard, and Ketterle}}]{shin2004distillation}
\bibinfo{author}{\bibfnamefont{Y.}~\bibnamefont{Shin}},
  \bibinfo{author}{\bibfnamefont{M.}~\bibnamefont{Saba}},
  \bibinfo{author}{\bibfnamefont{A.}~\bibnamefont{Schirotzek}},
  \bibinfo{author}{\bibfnamefont{T.}~\bibnamefont{Pasquini}},
  \bibinfo{author}{\bibfnamefont{A.}~\bibnamefont{Leanhardt}},
  \bibinfo{author}{\bibfnamefont{D.}~\bibnamefont{Pritchard}},
  \bibnamefont{and} \bibinfo{author}{\bibfnamefont{W.}~\bibnamefont{Ketterle}},
  \bibinfo{journal}{Physical Review Letters} \textbf{\bibinfo{volume}{92}},
  \bibinfo{pages}{150401} (\bibinfo{year}{2004}).

\bibitem[{\citenamefont{Potnis et~al.}(2017)\citenamefont{Potnis, Ramos, Maeda,
  Carr, and Steinberg}}]{potnis2017interaction}
\bibinfo{author}{\bibfnamefont{S.}~\bibnamefont{Potnis}},
  \bibinfo{author}{\bibfnamefont{R.}~\bibnamefont{Ramos}},
  \bibinfo{author}{\bibfnamefont{K.}~\bibnamefont{Maeda}},
  \bibinfo{author}{\bibfnamefont{L.~D.} \bibnamefont{Carr}}, \bibnamefont{and}
  \bibinfo{author}{\bibfnamefont{A.~M.} \bibnamefont{Steinberg}},
  \bibinfo{journal}{Physical review letters} \textbf{\bibinfo{volume}{118}},
  \bibinfo{pages}{060402} (\bibinfo{year}{2017}).

\bibitem[{\citenamefont{Cristofolini et~al.}(2012)\citenamefont{Cristofolini,
  Christmann, Tsintzos, Deligeorgis, Konstantinidis, Hatzopoulos, Savvidis, and
  Baumberg}}]{cristofolini2012coupling}
\bibinfo{author}{\bibfnamefont{P.}~\bibnamefont{Cristofolini}},
  \bibinfo{author}{\bibfnamefont{G.}~\bibnamefont{Christmann}},
  \bibinfo{author}{\bibfnamefont{S.~I.} \bibnamefont{Tsintzos}},
  \bibinfo{author}{\bibfnamefont{G.}~\bibnamefont{Deligeorgis}},
  \bibinfo{author}{\bibfnamefont{G.}~\bibnamefont{Konstantinidis}},
  \bibinfo{author}{\bibfnamefont{Z.}~\bibnamefont{Hatzopoulos}},
  \bibinfo{author}{\bibfnamefont{P.~G.} \bibnamefont{Savvidis}},
  \bibnamefont{and} \bibinfo{author}{\bibfnamefont{J.~J.}
  \bibnamefont{Baumberg}}, \bibinfo{journal}{Science}
  \textbf{\bibinfo{volume}{336}}, \bibinfo{pages}{704} (\bibinfo{year}{2012}).

\bibitem[{\citenamefont{Carusotto and Ciuti}(2013)}]{carusotto2013quantum}
\bibinfo{author}{\bibfnamefont{I.}~\bibnamefont{Carusotto}} \bibnamefont{and}
  \bibinfo{author}{\bibfnamefont{C.}~\bibnamefont{Ciuti}},
  \bibinfo{journal}{Reviews of Modern Physics} \textbf{\bibinfo{volume}{85}},
  \bibinfo{pages}{299} (\bibinfo{year}{2013}).

\bibitem[{\citenamefont{Grifoni and H{\"a}nggi}(1998)}]{grifoni1998driven}
\bibinfo{author}{\bibfnamefont{M.}~\bibnamefont{Grifoni}} \bibnamefont{and}
  \bibinfo{author}{\bibfnamefont{P.}~\bibnamefont{H{\"a}nggi}},
  \bibinfo{journal}{Physics Reports} \textbf{\bibinfo{volume}{304}},
  \bibinfo{pages}{229} (\bibinfo{year}{1998}).

\bibitem[{\citenamefont{Eckardt}(2017)}]{eckardt2017colloquium}
\bibinfo{author}{\bibfnamefont{A.}~\bibnamefont{Eckardt}},
  \bibinfo{journal}{Reviews of Modern Physics} \textbf{\bibinfo{volume}{89}},
  \bibinfo{pages}{011004} (\bibinfo{year}{2017}).

\bibitem[{\citenamefont{Arndt et~al.}(1999)\citenamefont{Arndt, Nairz,
  Vos-Andreae, Keller, Van~der Zouw, and Zeilinger}}]{arndt1999wave}
\bibinfo{author}{\bibfnamefont{M.}~\bibnamefont{Arndt}},
  \bibinfo{author}{\bibfnamefont{O.}~\bibnamefont{Nairz}},
  \bibinfo{author}{\bibfnamefont{J.}~\bibnamefont{Vos-Andreae}},
  \bibinfo{author}{\bibfnamefont{C.}~\bibnamefont{Keller}},
  \bibinfo{author}{\bibfnamefont{G.}~\bibnamefont{Van~der Zouw}},
  \bibnamefont{and}
  \bibinfo{author}{\bibfnamefont{A.}~\bibnamefont{Zeilinger}},
  \bibinfo{journal}{nature} \textbf{\bibinfo{volume}{401}},
  \bibinfo{pages}{680} (\bibinfo{year}{1999}).

\bibitem[{\citenamefont{Nimmrichter and
  Hornberger}(2013)}]{nimmrichter2013macroscopicity}
\bibinfo{author}{\bibfnamefont{S.}~\bibnamefont{Nimmrichter}} \bibnamefont{and}
  \bibinfo{author}{\bibfnamefont{K.}~\bibnamefont{Hornberger}},
  \bibinfo{journal}{Physical review letters} \textbf{\bibinfo{volume}{110}},
  \bibinfo{pages}{160403} (\bibinfo{year}{2013}).

\bibitem[{\citenamefont{Serwane et~al.}(2011)\citenamefont{Serwane, Z{\"u}rn,
  Lompe, Ottenstein, Wenz, and Jochim}}]{serwane2011deterministic}
\bibinfo{author}{\bibfnamefont{F.}~\bibnamefont{Serwane}},
  \bibinfo{author}{\bibfnamefont{G.}~\bibnamefont{Z{\"u}rn}},
  \bibinfo{author}{\bibfnamefont{T.}~\bibnamefont{Lompe}},
  \bibinfo{author}{\bibfnamefont{T.}~\bibnamefont{Ottenstein}},
  \bibinfo{author}{\bibfnamefont{A.}~\bibnamefont{Wenz}}, \bibnamefont{and}
  \bibinfo{author}{\bibfnamefont{S.}~\bibnamefont{Jochim}},
  \bibinfo{journal}{Science} \textbf{\bibinfo{volume}{332}},
  \bibinfo{pages}{336} (\bibinfo{year}{2011}).

\bibitem[{\citenamefont{Farhi et~al.}(2014)\citenamefont{Farhi, Goldstone, and
  Gutmann}}]{farhi2014quantum}
\bibinfo{author}{\bibfnamefont{E.}~\bibnamefont{Farhi}},
  \bibinfo{author}{\bibfnamefont{J.}~\bibnamefont{Goldstone}},
  \bibnamefont{and} \bibinfo{author}{\bibfnamefont{S.}~\bibnamefont{Gutmann}},
  \bibinfo{journal}{arXiv preprint arXiv:1411.4028}  (\bibinfo{year}{2014}).

\bibitem[{\citenamefont{Albash and Lidar}(2018)}]{albash2018adiabatic}
\bibinfo{author}{\bibfnamefont{T.}~\bibnamefont{Albash}} \bibnamefont{and}
  \bibinfo{author}{\bibfnamefont{D.~A.} \bibnamefont{Lidar}},
  \bibinfo{journal}{Reviews of Modern Physics} \textbf{\bibinfo{volume}{90}},
  \bibinfo{pages}{015002} (\bibinfo{year}{2018}).

\bibitem[{\citenamefont{Altshuler et~al.}(2010)\citenamefont{Altshuler, Krovi,
  and Roland}}]{altshulerkrovi2010}
\bibinfo{author}{\bibfnamefont{B.}~\bibnamefont{Altshuler}},
  \bibinfo{author}{\bibfnamefont{H.}~\bibnamefont{Krovi}}, \bibnamefont{and}
  \bibinfo{author}{\bibfnamefont{J.}~\bibnamefont{Roland}},
  \bibinfo{journal}{Proceedings of the National Academy of Sciences}
  \textbf{\bibinfo{volume}{107}}, \bibinfo{pages}{12446}
  (\bibinfo{year}{2010}).

\bibitem[{\citenamefont{Knysh}(2016)}]{knysh2016}
\bibinfo{author}{\bibfnamefont{S.}~\bibnamefont{Knysh}},
  \bibinfo{journal}{Nature communications} \textbf{\bibinfo{volume}{7}}
  (\bibinfo{year}{2016}).

\bibitem[{\citenamefont{Tang and Kapit}(2021)}]{tang2021unconventional}
\bibinfo{author}{\bibfnamefont{Z.}~\bibnamefont{Tang}} \bibnamefont{and}
  \bibinfo{author}{\bibfnamefont{E.}~\bibnamefont{Kapit}},
  \bibinfo{journal}{Physical Review A} \textbf{\bibinfo{volume}{103}},
  \bibinfo{pages}{032612} (\bibinfo{year}{2021}).

\bibitem[{\citenamefont{Kapit and Oganesyan}(2021)}]{kapit2021noise}
\bibinfo{author}{\bibfnamefont{E.}~\bibnamefont{Kapit}} \bibnamefont{and}
  \bibinfo{author}{\bibfnamefont{V.}~\bibnamefont{Oganesyan}},
  \bibinfo{journal}{Quantum Science and Technology}
  \textbf{\bibinfo{volume}{6}}, \bibinfo{pages}{025013} (\bibinfo{year}{2021}).

\bibitem[{\citenamefont{Oka and Kitamura}(2019)}]{oka2019floquet}
\bibinfo{author}{\bibfnamefont{T.}~\bibnamefont{Oka}} \bibnamefont{and}
  \bibinfo{author}{\bibfnamefont{S.}~\bibnamefont{Kitamura}},
  \bibinfo{journal}{Annual Review of Condensed Matter Physics}
  \textbf{\bibinfo{volume}{10}}, \bibinfo{pages}{387} (\bibinfo{year}{2019}).

\bibitem[{\citenamefont{Trotter}(1959)}]{trotter1959product}
\bibinfo{author}{\bibfnamefont{H.~F.} \bibnamefont{Trotter}},
  \bibinfo{journal}{Proceedings of the American Mathematical Society}
  \textbf{\bibinfo{volume}{10}}, \bibinfo{pages}{545} (\bibinfo{year}{1959}).

\bibitem[{\citenamefont{Schmitt et~al.}(2022)\citenamefont{Schmitt, Rams,
  Dziarmaga, Heyl, and Zurek}}]{schmitt2022quantum}
\bibinfo{author}{\bibfnamefont{M.}~\bibnamefont{Schmitt}},
  \bibinfo{author}{\bibfnamefont{M.~M.} \bibnamefont{Rams}},
  \bibinfo{author}{\bibfnamefont{J.}~\bibnamefont{Dziarmaga}},
  \bibinfo{author}{\bibfnamefont{M.}~\bibnamefont{Heyl}}, \bibnamefont{and}
  \bibinfo{author}{\bibfnamefont{W.~H.} \bibnamefont{Zurek}},
  \bibinfo{journal}{Science Advances} \textbf{\bibinfo{volume}{8}},
  \bibinfo{pages}{eabl6850} (\bibinfo{year}{2022}).

\bibitem[{\citenamefont{Bukov et~al.}(2016)\citenamefont{Bukov, Kolodrubetz,
  and Polkovnikov}}]{bukov2016schrieffer}
\bibinfo{author}{\bibfnamefont{M.}~\bibnamefont{Bukov}},
  \bibinfo{author}{\bibfnamefont{M.}~\bibnamefont{Kolodrubetz}},
  \bibnamefont{and}
  \bibinfo{author}{\bibfnamefont{A.}~\bibnamefont{Polkovnikov}},
  \bibinfo{journal}{Physical review letters} \textbf{\bibinfo{volume}{116}},
  \bibinfo{pages}{125301} (\bibinfo{year}{2016}).

\bibitem[{\citenamefont{Rudner and Lindner}(2020)}]{rudner2020floquet}
\bibinfo{author}{\bibfnamefont{M.~S.} \bibnamefont{Rudner}} \bibnamefont{and}
  \bibinfo{author}{\bibfnamefont{N.~H.} \bibnamefont{Lindner}},
  \bibinfo{journal}{arXiv preprint arXiv:2003.08252}  (\bibinfo{year}{2020}).

\bibitem[{\citenamefont{Abanin et~al.}(2015)\citenamefont{Abanin, De~Roeck, and
  Huveneers}}]{abanin2015exponentially}
\bibinfo{author}{\bibfnamefont{D.~A.} \bibnamefont{Abanin}},
  \bibinfo{author}{\bibfnamefont{W.}~\bibnamefont{De~Roeck}}, \bibnamefont{and}
  \bibinfo{author}{\bibfnamefont{F.}~\bibnamefont{Huveneers}},
  \bibinfo{journal}{Physical review letters} \textbf{\bibinfo{volume}{115}},
  \bibinfo{pages}{256803} (\bibinfo{year}{2015}).

\bibitem[{\citenamefont{Mallayya and Rigol}(2019)}]{mallayya2019heating}
\bibinfo{author}{\bibfnamefont{K.}~\bibnamefont{Mallayya}} \bibnamefont{and}
  \bibinfo{author}{\bibfnamefont{M.}~\bibnamefont{Rigol}},
  \bibinfo{journal}{Physical review letters} \textbf{\bibinfo{volume}{123}},
  \bibinfo{pages}{240603} (\bibinfo{year}{2019}).

\bibitem[{\citenamefont{Else et~al.}(2020)\citenamefont{Else, Ho, and
  Dumitrescu}}]{else2020long}
\bibinfo{author}{\bibfnamefont{D.~V.} \bibnamefont{Else}},
  \bibinfo{author}{\bibfnamefont{W.~W.} \bibnamefont{Ho}}, \bibnamefont{and}
  \bibinfo{author}{\bibfnamefont{P.~T.} \bibnamefont{Dumitrescu}},
  \bibinfo{journal}{Physical Review X} \textbf{\bibinfo{volume}{10}},
  \bibinfo{pages}{021032} (\bibinfo{year}{2020}).

\bibitem[{\citenamefont{Machado et~al.}(2020)\citenamefont{Machado, Else,
  Kahanamoku-Meyer, Nayak, and Yao}}]{machado2020long}
\bibinfo{author}{\bibfnamefont{F.}~\bibnamefont{Machado}},
  \bibinfo{author}{\bibfnamefont{D.~V.} \bibnamefont{Else}},
  \bibinfo{author}{\bibfnamefont{G.~D.} \bibnamefont{Kahanamoku-Meyer}},
  \bibinfo{author}{\bibfnamefont{C.}~\bibnamefont{Nayak}}, \bibnamefont{and}
  \bibinfo{author}{\bibfnamefont{N.~Y.} \bibnamefont{Yao}},
  \bibinfo{journal}{Physical Review X} \textbf{\bibinfo{volume}{10}},
  \bibinfo{pages}{011043} (\bibinfo{year}{2020}).

\bibitem[{\citenamefont{Zhao et~al.}(2021)\citenamefont{Zhao, Mintert,
  Moessner, and Knolle}}]{zhao2021random}
\bibinfo{author}{\bibfnamefont{H.}~\bibnamefont{Zhao}},
  \bibinfo{author}{\bibfnamefont{F.}~\bibnamefont{Mintert}},
  \bibinfo{author}{\bibfnamefont{R.}~\bibnamefont{Moessner}}, \bibnamefont{and}
  \bibinfo{author}{\bibfnamefont{J.}~\bibnamefont{Knolle}},
  \bibinfo{journal}{Physical Review Letters} \textbf{\bibinfo{volume}{126}},
  \bibinfo{pages}{040601} (\bibinfo{year}{2021}).

\bibitem[{\citenamefont{Shkedrov et~al.}(2022)\citenamefont{Shkedrov, Menashes,
  Ness, Vainbaum, Altman, and Sagi}}]{shkedrov2022absence}
\bibinfo{author}{\bibfnamefont{C.}~\bibnamefont{Shkedrov}},
  \bibinfo{author}{\bibfnamefont{M.}~\bibnamefont{Menashes}},
  \bibinfo{author}{\bibfnamefont{G.}~\bibnamefont{Ness}},
  \bibinfo{author}{\bibfnamefont{A.}~\bibnamefont{Vainbaum}},
  \bibinfo{author}{\bibfnamefont{E.}~\bibnamefont{Altman}}, \bibnamefont{and}
  \bibinfo{author}{\bibfnamefont{Y.}~\bibnamefont{Sagi}},
  \bibinfo{journal}{Physical Review X} \textbf{\bibinfo{volume}{12}},
  \bibinfo{pages}{011041} (\bibinfo{year}{2022}).

\bibitem[{\citenamefont{Ikeda et~al.}(2023)\citenamefont{Ikeda, Sugiura, and
  Polkovnikov}}]{ikeda2023robust}
\bibinfo{author}{\bibfnamefont{T.~N.} \bibnamefont{Ikeda}},
  \bibinfo{author}{\bibfnamefont{S.}~\bibnamefont{Sugiura}}, \bibnamefont{and}
  \bibinfo{author}{\bibfnamefont{A.}~\bibnamefont{Polkovnikov}},
  \bibinfo{journal}{arXiv preprint arXiv:2311.16217}  (\bibinfo{year}{2023}).

\bibitem[{\citenamefont{Kim et~al.}(2023)\citenamefont{Kim, Eddins, Anand, Wei,
  Van Den~Berg, Rosenblatt, Nayfeh, Wu, Zaletel, Temme
  et~al.}}]{kim2023evidence}
\bibinfo{author}{\bibfnamefont{Y.}~\bibnamefont{Kim}},
  \bibinfo{author}{\bibfnamefont{A.}~\bibnamefont{Eddins}},
  \bibinfo{author}{\bibfnamefont{S.}~\bibnamefont{Anand}},
  \bibinfo{author}{\bibfnamefont{K.~X.} \bibnamefont{Wei}},
  \bibinfo{author}{\bibfnamefont{E.}~\bibnamefont{Van Den~Berg}},
  \bibinfo{author}{\bibfnamefont{S.}~\bibnamefont{Rosenblatt}},
  \bibinfo{author}{\bibfnamefont{H.}~\bibnamefont{Nayfeh}},
  \bibinfo{author}{\bibfnamefont{Y.}~\bibnamefont{Wu}},
  \bibinfo{author}{\bibfnamefont{M.}~\bibnamefont{Zaletel}},
  \bibinfo{author}{\bibfnamefont{K.}~\bibnamefont{Temme}},
  \bibnamefont{et~al.}, \bibinfo{journal}{Nature}
  \textbf{\bibinfo{volume}{618}}, \bibinfo{pages}{500} (\bibinfo{year}{2023}).

\bibitem[{\citenamefont{Eckstein et~al.}(2023)\citenamefont{Eckstein,
  Mansuroglu, Czarnik, Zhu, Hartmann, Cincio, Sornborger, and
  Holmes}}]{eckstein2023large}
\bibinfo{author}{\bibfnamefont{T.}~\bibnamefont{Eckstein}},
  \bibinfo{author}{\bibfnamefont{R.}~\bibnamefont{Mansuroglu}},
  \bibinfo{author}{\bibfnamefont{P.}~\bibnamefont{Czarnik}},
  \bibinfo{author}{\bibfnamefont{J.-X.} \bibnamefont{Zhu}},
  \bibinfo{author}{\bibfnamefont{M.~J.} \bibnamefont{Hartmann}},
  \bibinfo{author}{\bibfnamefont{L.}~\bibnamefont{Cincio}},
  \bibinfo{author}{\bibfnamefont{A.~T.} \bibnamefont{Sornborger}},
  \bibnamefont{and} \bibinfo{author}{\bibfnamefont{Z.}~\bibnamefont{Holmes}},
  \bibinfo{journal}{arXiv preprint arXiv:2303.02209}  (\bibinfo{year}{2023}).

\end{thebibliography}
\end{document}


\newcommand{\of}[1]{\left( #1 \right)}
\newcommand{\sqof}[1]{\left[ #1 \right]}
\newcommand{\abs}[1]{\left| #1 \right|}
\newcommand{\avg}[1]{\left< #1 \right>}
\newcommand{\cuof}[1]{\left \{ #1 \right \} }
\newcommand{\bra}[1]{\left < #1 \right | }
\newcommand{\ket}[1]{\left | #1 \right > }
\newcommand{\pil}{\frac{\pi}{L}}
\newcommand{\bx}{\mathbf{x}}
\newcommand{\by}{\mathbf{y}}
\newcommand{\bk}{\mathbf{k}}
\newcommand{\bp}{\mathbf{p}}
\newcommand{\bl}{\mathbf{l}}
\newcommand{\bq}{\mathbf{q}}
\newcommand{\bs}{\mathbf{s}}
\newcommand{\psibar}{\overline{\psi}}
\newcommand{\svec}{\overrightarrow{\sigma}}
\newcommand{\dvec}{\overrightarrow{\partial}}
\newcommand{\bA}{\mathbf{A}}
\newcommand{\bdelta}{\mathbf{\delta}}
\newcommand{\bK}{\mathbf{K}}
\newcommand{\bQ}{\mathbf{Q}}
\newcommand{\bG}{\mathbf{G}}
\newcommand{\bw}{\mathbf{w}}
\newcommand{\bL}{\mathbf{L}}
\newcommand{\ohat}{\widehat{O}}
\newcommand{\up}{\uparrow}
\newcommand{\down}{\downarrow}
\newcommand{\MM}{\mathcal{M}}
\newcommand{\MN}{\mathcal{N}}
\newcommand{\MR}{\mathcal{R}}
\newcommand{\tW}{\tilde{W}}
\newcommand{\tX}{\tilde{X}}
\newcommand{\tY}{\tilde{Y}}
\newcommand{\tZ}{\tilde{Z}}
\newcommand{\tOm}{\tilde{\Omega}}
\newcommand{\barA}{\bar{\alpha}}

\author{George Grattan$^{1,2}$, Brandon A. Barton$^{1,3}$, Sean Feeney$^{1}$, Gianni Mossi$^{5,6}$, Pratik Patnaik$^{1,4}$, Jacob C. Sagal$^{1}$, Lincoln D. Carr$^{1,3,4}$, Vadim Oganesyan$^{7,8,9}$, and Eliot Kapit$^{1,4 *}$}
\address{$^1$ Quantum Engineering Program, Colorado School of Mines, 1523 Illinois St, Golden CO 80401}
\address{$^2$ Department of Computer Science, Colorado School of Mines, 1500 Illinois St, Golden CO 80401}
\address{$^3$ Department of Applied Mathematics and Statistics, Colorado School of Mines, 1500 Illinois St, Golden CO 80401}
\address{$^4$ Department of Physics, Colorado School of Mines, 1523 Illinois St, Golden CO 80401}

\address{$^5$ KBR, Inc., 601 Jefferson St., Houston, TX 77002, USA.}
\address{$^6$ Quantum Artificial Intelligence Lab. (QuAIL), NASA Ames Research Center, Moffett Field, CA 94035, USA.}
\address{$^7$ Department of Physics and Astronomy, College of Staten Island, CUNY, Staten Island, NY 10314, USA}
\address{$^8$ Physics program and Initiative for the Theoretical Sciences, The Graduate Center, CUNY, New York, NY 10016, USA}
\address{$^9$ Center for Computational Quantum Physics, Flatiron Institute, 162 5th Avenue, New York, NY 10010, USA}

\email{ekapit@mines.edu}

\title{Supplemental material: Exponential acceleration of macroscopic quantum tunneling in a Floquet Ising model}

\maketitle

The content of this supplement is organized into a presentation of additional details on simulation in main text (Sec. \ref{sec:methods}), high orders resummation of tunnelling processes in the ordinary static TFIM that is required to reproduce the correct dependence of the tunnelling rate in the ferromagnetic phase and,  to the best of our knowledge unavailable in existing literature (Sec. \ref{sec:FA}), an extension of this theory to the effective Hamiltonian generated by periodic driving (Sec. \ref{sec:FAFloq}), the analytic perturbative and numeric non-perturbative calculation of the effective Hamiltonian (Sec. \ref{sec:Heff}) which is non-integrable even in one dimension. , Section \ref{sec:floqadia}
takes a closer look at the dynamics and its reduced spectral representation in terms of just two eigenstates of the Floquet unitary. Here, we clearly demonstrate the importance of adiabaticity and intra-period (so called micromotion) to our findings. In closing Section \ref{sec:taveraging} we explore how our results could appear in a measurement apparatus that performs time-averaging, as opposed to stroboscopic sampling.

Much of this discussion focuses primarily on TFIM chains, we expect the physical picture to persist to 2D (as supported by the numerical results presented in main text).

\subsection{Simulation details\label{sec:methods}}
We now present some more precise details of our simulations. Unless otherwise noted, we used a ramp time $T_r = N/8$ with a smooth sinusoidal protocol. To avoid edge effects, we used periodic boundary conditions in all cases. During evolution we measured the ferromagnetic order parameter $\avg{Z_i Z_j}$ between the two most distant qubits, to characterize the instantaneous state of the system.This quantity was time-averaged over the plateau to eliminate high frequency modulations. To simulate evolution under a high frequency AC drive, we chose a decreasing timestep $dt = c/f$ sufficiently small to appropriately sample it. $dt$ too large leads to severe Trotter error and nonsensical output; $c$ between 0.15 and 0.4 was sufficient for faithful simulations, with smaller values necessary when $\alpha$ is large. Our results thus simulate continuous time and are \emph{not} due to time discretization. The Trotter error was eliminated by choosing an appropriate timestep $dt = c/f$ with value of $c$ between 0.25 and 0.125. The bond dimension $\chi$ values in our 1d TEBD simulations was chosen up to $50$. 

For a given system size $N$ and $\alpha/f$ value, the Rabi frequency $\Omega_R$ is estimated by simulating the Floquet protocol for times $t_1, t_2, \ldots, t_n$ obtaining the tunnelling probabilities $p_1, p_2, \ldots, p_n$, where $p_i \equiv \lvert \langle 1\cdots 1 \rvert U(t_i) \lvert 0\cdots 0 \rangle \rvert^2$. $\Omega_R$ is then extracted by fitting the data with the function $p(t) = \sin^2(\Omega_R t + \phi_{init})$, as shown in Fig. 1(right panel). For all simulations, the ramptimes and plateau times were chosen so that (i) heating observed would introduce only negligible effects in the extracted value of the Rabi frequency, and (ii) the probed times would allow us to follow at least one oscillation of the tunnelling probability. 

When fitting the data, we used a Fast Fourier Transform on the state population data to obtain a warm start for the frequency. Additionally we fit the Rabi frequency for both $|\langle \psi(T) | 1 \rangle|^2$ and $|\langle \psi(T) | 0 \rangle |^2$ and the reported value is the mean of these two values. There were instances where we needed to add a phase of $\pi/2$ to $\phi_{init}$ in order to help the model to fit the data, this is likely due to the size dependent simulation times, which caused inconsistent "initial" populations/ initial phases. Additionally in instances with a heating we used a model that included a decaying exponent term, namely $p(t) = e^{-b t}\sin^2(\Omega_R t + \phi_{init})$ and we found little inconsistency in the extracted values for $\Omega_R$ in these cases. We then logarithmically scaled the values of $\Omega_R$, as seen in Supplementary Figure  \ref{fig:2d_Rabi_scaling_qsim} and fit them to $\log(\Omega_R) = A + \Upsilon*N$ in order to extract the difficulty exponent $\Upsilon$ seen in 3.
In 1d our fits were to $\log(\Omega_R) = A + B \log N +  \Upsilon*N$, as the larger system sizes accessible through TEBD allowed us to resolve polynomial prefactors more accurately. In both cases we observe that the Rabi frequency decays exponentially with the system size at small values of $\alpha/f$ and enters a ``fast tunnelling'' regime at larger values. In the 1d case small positive values of the difficulty exponent $\Upsilon$ (see Fig.~\ref{fig:1d_notable_parameters_TEBD}) are observed in the fast tunnelling regime while $\Omega_R$ is still slowly decaying with the system size. We believe these to be an artifact of the fit due to finite size, and interpret them to be compatible with $\Upsilon=0$.

  \begin{figure}

     \includegraphics[width=0.49\textwidth]{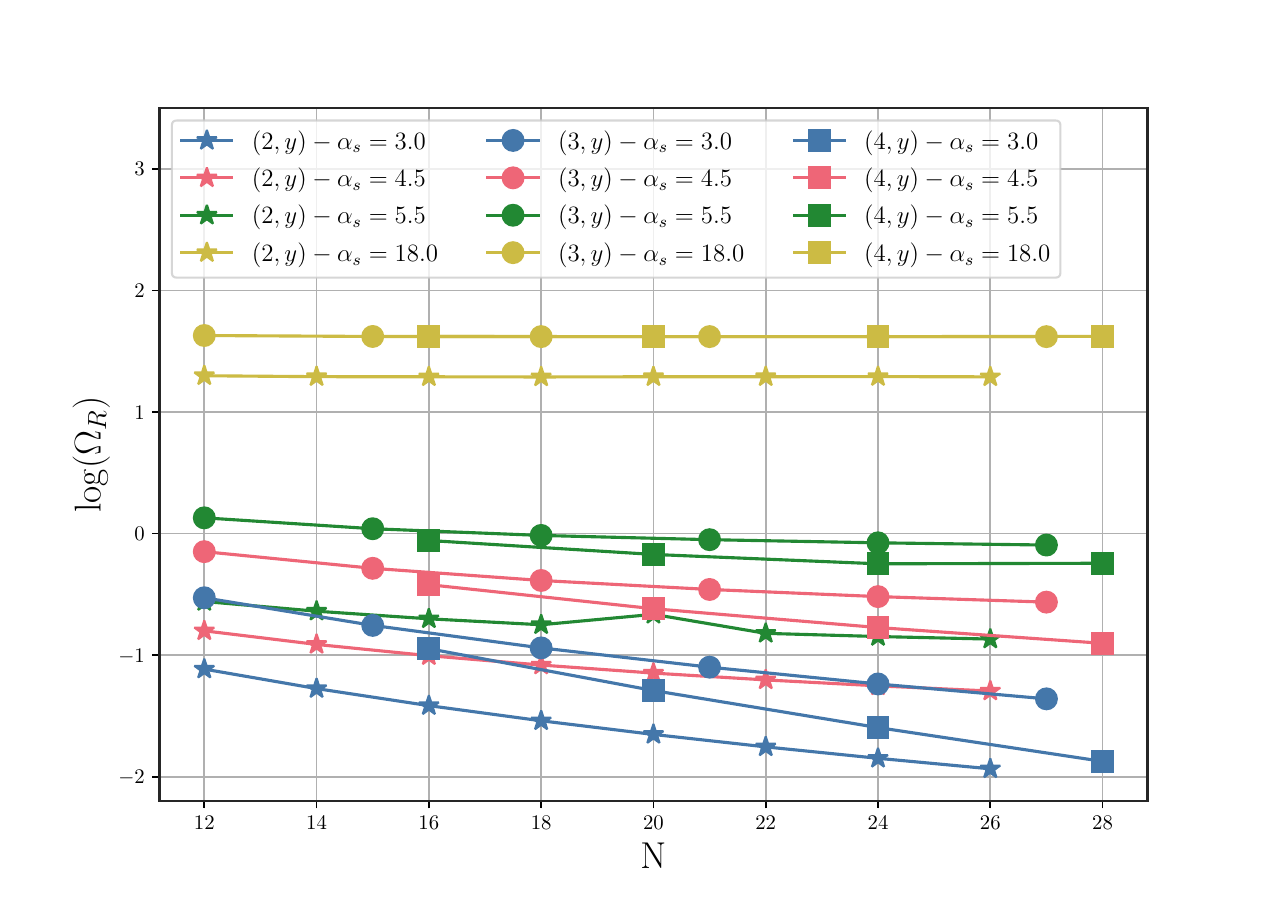}
     \caption{Lograthmically scaled Rabi oscillations in the two-dimensional transverse-field model, with $\kappa = 2.5$. Collections of data such as this one are used to extract the difficulty scaling exponent and characterize the three drive strength regimes.}
     \label{fig:2d_Rabi_scaling_qsim}
 \end{figure}

\begin{figure}

     \includegraphics[width=0.45\textwidth]{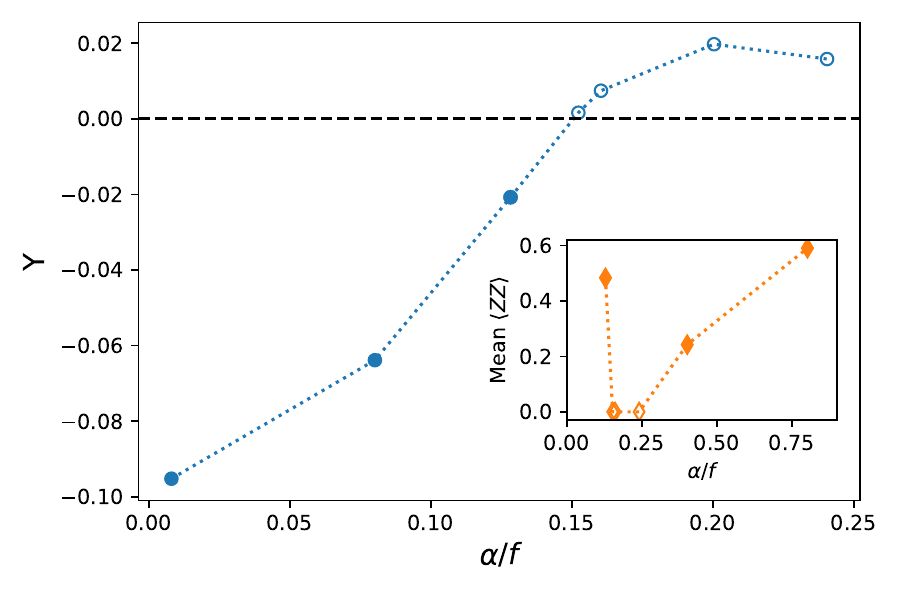}
     \caption{Difficulty exponents $\Upsilon$ extracted from Fig. 2.
     Solid dots indicate datapoints where $\Omega_R$ decays exponentially, while empty circles are compatible with subexponential decay. Inset: time-averaged values for the two-point correlator from Fig. 2.
     Filled diamonds indicate values of $\alpha/f$ where we observe saturation of $\langle Z_0Z_{L/2} \rangle$ in the range of system sizes observed. Empty diamonds label cases where no such saturation is observed, which we conservatively interpret a zero magnetization.}
     \label{fig:1d_notable_parameters_TEBD}
 \end{figure}

\subsection{Theory of MQT in the 1d TFIM -- all orders analysis}
\label{sec:FA}
We consider the 1d transverse field Ising model, in the ferromagnetic phase. Our Hamiltonian is
\begin{equation}
H = - \sum_j \of{ \kappa X_j + J Z_j Z_{j+1} }
\end{equation}
We assume periodic boundary conditions, for simplicity. Our goal is to compute the exponentially small tunneling rate $\Omega_0$ between degenerate ground states in the ferromagnetic phase, where $\kappa < J$, through $L$th order perturbation theory in $\kappa$ using the Ising Hamiltonian as the base Hamiltonian. Fundamentally, this is a sum of $L!$ flip sequences connecting the two classical ground states \cite{bravyi2011schrieffer,pietracaprina2016forward,baldwinlaumann2016,baldwinlaumann2017,scardicchio2017perturbation,baldwin2018quantum}, e.g.
\begin{equation}
\Omega_0 \simeq \kappa^L \sum_{perm. seq \cuof{s_j} } \prod_{k=1}^{L} \frac{1}{E_{\cuof{s_1,s_2... s_k}}}
\end{equation}

Now, with $L!$ terms and strong path dependence in the denominator, this expression looks hopeless. But it turns out there are two limits where one can evaluate sums of this form, to good approximation. One is infinite-dimensional (e.g. random) graphs or hypergraphs, where the influence of specific sequences with (comparatively) low intermediate energies, which have higher weight in the sum, is swamped by the combinatorical proliferation of random sequences, an issue we take up in a forthcoming publication. The other is limit is 1d, e.g. this problem. It will become clear from the structure of the derivation why 2d is considerably harder and we do not attempt it here.

The trick to evaluating this sum is to organize the sets of terms by the maximum number of domain wall pairs created at intermediate steps. We'll start with the simplest process, where there is just a single pair of domain walls at all steps. This process begins by enacting a single flip at any of the $L$ sites, which has an energy cost $4J$. The next flip is at one of the two sites adjacent to this, which moves one of the two domain walls further apart by one step. The intermediate energy is still $4J$, but there are two choices we can make, which gives us a combinatorical factor of 2 in the numerator. The same goes for the third flip, fourth flip, and so on, ``unzipping" down the chain and accumulating a factor of $\kappa/2J$ at each step. Noting that there were $L$ choices for the first site to flip, 
\begin{eqnarray}\label{onepair}
\Omega_0^{(1)} = L \frac{\kappa^2}{4J} \of{\frac{\kappa}{2J}}^{L-2} = L \frac{\kappa}{2} \of{ \frac{\kappa}{2J}}^{L-1}
\end{eqnarray}
This is the lowest energy path possible, so one would be tempted to stop here. But doing so underestimates $\Omega_0$ by a factor of $2^L$, which isn't great. To get the correct value--or at least, the correct exponent--we need to consider higher order terms.

Let's now consider the set of all processes where a second pair of domain walls is is created at some intermediate stage. Assume that $k=1$ is the first step and the first $k$ flips simply create the first pair and separate them. Then we now have a total of $\of{L-k-2}$ choices for where to nucleate the second pair (the $-2$ comes from two sites just moving existing domain walls around), and the energy cost once this is done is $8J$ (compared to $1/2J$, counting the combinatorics, for a flip at this stage that just moves a domain wall). This is the energy cost that now shows up in the dominator at each subsequent order until a pair of domain walls eventually fuses, but we now have 4 sites we can flip to move a domain wall around, so we get a factor of $\kappa/2J$ at each subsequent step just like if we had a single pair, and once one of the pairs is eliminated we're back to the same result as above. Combining all these factors, and summing over space and ``time" points where the second pair is created, we get:
\begin{eqnarray}
\Omega_0^{(2)} \simeq \sum_{k=1}^{L-2} \frac{L-k-2}{4} \Omega_0^{(1)}.
\end{eqnarray} 

Now with three intermediate pairs. The third pair is nucleated at any step after $k$, called step $l$. At this step there are $\of{L-l-4}$ sites to choose from, and the energy cost is $12J$ (corresponding to a factor of 1/6 in comparison to a flip that just moves a domain wall). Noting the factor of $1/4$ from the previous sequence, and ignoring the combinatorically subleading cases where the second pair of domain walls was destroyed before this third pair was created, we find
\begin{eqnarray}
\Omega_0^{(3)} \simeq \sum_{k=1}^{L-2} \frac{L-k-2}{4} \sum_{l=k}^{L-4} \frac{L-l-4}{6} \Omega_0^{(1)}.
\end{eqnarray} 
We can generalize this to four intermediate pairs, by inspection:
\begin{eqnarray}
\Omega_0^{(4)} \simeq \sum_{k=1}^{L-2} \frac{L-k-2}{4} \sum_{l=k}^{L-4} \frac{L-l-4}{6} \sum_{m=l}^{L-4} \frac{L-m-6}{8} \Omega_0^{(1)}.
\end{eqnarray}
If we evaluate these sums in mathematica, a pattern quickly emerges, and a bit of algebra and inspection shows that at order $p$ we have
\begin{eqnarray}
\Omega_0^{(p)} \simeq \sqof{ \frac{p}{2^{2p-1} \of{p!}^2 } \prod_{k=2}^{2p-1} \of{L - k} } \Omega_0^{(1)}.
\end{eqnarray}
The product evaluates to a Pochhammer function. Our total tunneling rate is given by taking this sum out to $L/2$, the maximum number of domain wall pairs possible in this system:
\begin{eqnarray}\label{Om0full}
\Omega_0 \simeq L \frac{\kappa}{2} \of{ \frac{\kappa}{2J}}^{L-1} \sum_{p=1}^{L/2} \sqof{ \frac{p}{2^{2p-1} \of{p!}^2 } \prod_{k=2}^{2p-1} \of{L - k} }.
\end{eqnarray}
This is a sum Mathematica can do; we get
\begin{eqnarray}
\Omega_0 \simeq \frac{L}{4} \kappa \of{\frac{\kappa}{J} }^{L-1} \frac{\Gamma \of{L - 1/2 } }{\sqrt{\pi} L!}
\end{eqnarray}
Aymptotically, the ratio $\Gamma \of{L-1/2} / L!$ converges to $L^{-3/2}$, so up to an overall constant prefactor
\begin{eqnarray}\label{1dpred}
\Omega_0 \propto \frac{\kappa}{\sqrt{L}} \of{\frac{\kappa}{J} }^{L-1}.
\end{eqnarray}
This matches numerics extremely well, though numerical data suggests a polynomial prefactor of $L^{-\kappa/J}$ is a slightly better fit. That the exponent is such a close match is a real triumph for high order perturbation theory.

\subsection{Collective tunneling in the AC driven case}
\label{sec:FAFloq}
In the weak and crossover regimes, the average Hamiltonian formulation predicts both the renormalization of $\kappa$ and $J$ from fast oscillations, and the generation of new pair-flip terms that constructively interfere to accelerate tunneling (see next Sec. \ref{sec:Heff}). We can incorporate all these effects into the derivation of Eq.~\ref{1dpred} to predict both the reduction of the scaling exponent $\Upsilon$ in the weak driving regime, and the approximate location of the crossover to polynomial scaling by identifying where $\Upsilon \to 0$. The renormalization of $\kappa$ and $J$ is straightforward--following the arguments in \cite{kapit2022small}, $\kappa$ is reduced by a factor of $J_0 \of{2\alpha/f}$, where $J_0$ is a Bessel function. Note that the argument is $\alpha/f$ and not $\alpha/\omega$, respecting the convention stated in the main text of including a factor of $2 \pi$ in front of $H \of{t}$ in time evolution. Since the ferromagnetic coupling $J$ involves two-spin terms it is reduced by a factor of $J_0 \of{2 \alpha/f}^2$, at least in the limit of $\alpha/f$ small, and thus the ratio $\kappa/J$ is \emph{increased} by a factor of $1/J_0 \of{2 \alpha/f}$.

To incorporate the pair flip terms, we need to take into account two sets of processes. The first moving a domain wall by two steps, with matrix element $- J \frac{\alpha^2}{f^2}$ (as compared to moving it a single step with matrix $-\kappa$ in the DC transverse field), and creating a domain wall pair two sites apart with matrix element $- J \frac{\alpha^2}{f^2}$. The energy denominators depend on the Ising Hamiltonian and are thus unchanged beyond the renormalization of $J$ already discussed. We can incorporate these terms into the analysis leading to~\ref{1dpred}, at least approximately, by making two substitutions. First, noting that ``most" of the spin flips that contribute to the tunneling amplitude consist of moving domain walls, we can now replace any single flip (which contributes $\kappa/\of{2J}$ when all combinatorics are considered) with a pair flip with amplitude $2 \frac{\alpha^2}{f^2}$ (including $1/J$ for the energy denominator). This replaces two orders in the DC calculation, and so the combination of all such processes multiplies the tunneling rate by $\of{ 1 + 2 \frac{\alpha^2 J^2}{f^2 \kappa^2}}^L$, to decent approximation. The second modification is that every step that nucleates a new domain wall pair can be replaced by the corresponding process that creates such a pair two flips apart, which removes a factor of $\kappa/\of{4J}$ from the overall tunneling rate since it flips two spins instead of one. Combining all these contributions, we derive an AC-modified tunnel splitting
\begin{eqnarray}\label{OmAC}
\Omega_{R} \of{\kappa,\alpha,f, L} \simeq L \frac{\kappa}{2} \of{ \frac{\kappa}{2J \times J_0 \of{2 \frac{\alpha}{f}}}}^{L-1} \times \\
\of{ 1 + 2 \frac{\alpha^2 J^2}{f^2 \kappa^2}}^L \times \sum_{p=1}^{L/2} \sqof{ \frac{p \of{ 1 + 4 \frac{\alpha^2 J^2}{f^2 \kappa^2}}^p}{2^{2p-1} \of{p!}^2 } \prod_{k=2}^{2p-1} \of{L - k} }. \nonumber
\end{eqnarray}

This expression must be evaluated numerically, but it's straightforward to do so. The decay exponent decreases continuously with $\alpha/f$, and if this ratio becomes large enough the decay exponent crosses zero. At this point perturbation theory has broken down and we expect a crossover to polynomial decay (as observed in our simulations), though we cannot predict the degree of that polynomial with these methods. For $\kappa = 0.9$, Eq.~\ref{OmAC} predicts $\Upsilon \to 0$ when $\alpha/f \simeq 0.15$, in excellent agreement with the simulations shown in Fig. 2.

\subsection{Derivation of the effective Hamiltonian}
\label{sec:Heff}
We now turn to the calculation of $H_{eff}\equiv \log U/(i P)$, where the single period unitary is $U = {\cal T} e^{i \int_0^P dt H(t)}$. We will only be concerned with the structure of this operator away from ramps, i.e. when the problem is strictly periodic, i.e.
(recall Eq. 1)
\begin{align}
    &H(t) = \sum_{j} \kappa X_i+J Z_j Z_{j+1} 
    +\alpha \sin{{\frac{2\pi t}{P}}}\  Y_j,
    \label{eq:Hoft}
\end{align}
with Floquet period $P=(6\log L)^{-1}$ (and $J<0$).

Standard Floquet-Magnus perturbation theory, a sort of a short-time (high frequency) expansion, may be used to calculate $H_{eff}$ perturbatively (i.e. when $U$ is close to being an identity), with third-order triple commutator producing the leading nontrivial corrections to the first order "average Hamiltonian" terms (first two terms in the Eq. \ref{eq:Hoft})
\begin{align}
   & H^{(3)}_{eff}=
        \frac{1}{6}\int_0^P dt_1 \int_0^{t_1} dt_2 \int_0^{t_2} dt_3  
    \nonumber \\
    &
([H(t_1),[H(t_2),H(t_3)]    +[H(t_3),[H(t_2),H(t_1)]\\
&H_{eff}= \sum_{j} \kappa X_i-J Z_j Z_{j+1} 
\nonumber \\
     &+\frac{3 \alpha^2 P^2}{16 \pi^2} \sum_j 2 J \ ({\bf Z}_j {\bf Z}_{j+1}-{\bf X}_j {\bf X}_{j+1})-\kappa {\bf X}_j +\ldots.
\label{eq:magnus}
\end{align}
The leading correction terms both suppress static $ZZ$ and $X$ interactions and promote domain wall creation and hopping, in pairs via the $XX$ term; the $XX$ term breaks integrability of the 1D TFIM and would be expected to produce thermalization generically (but incorrectly, in this context -- see below);  the leading terms are strictly perturbative in $\alpha, \kappa, J$. Since we are actually working in the regime of large $\kappa\lesssim |J|=1$, we now turn to complementary non-perturbative derivation of $H_{eff}$.

\begin{figure}
    \centering
    \includegraphics[width=\linewidth]{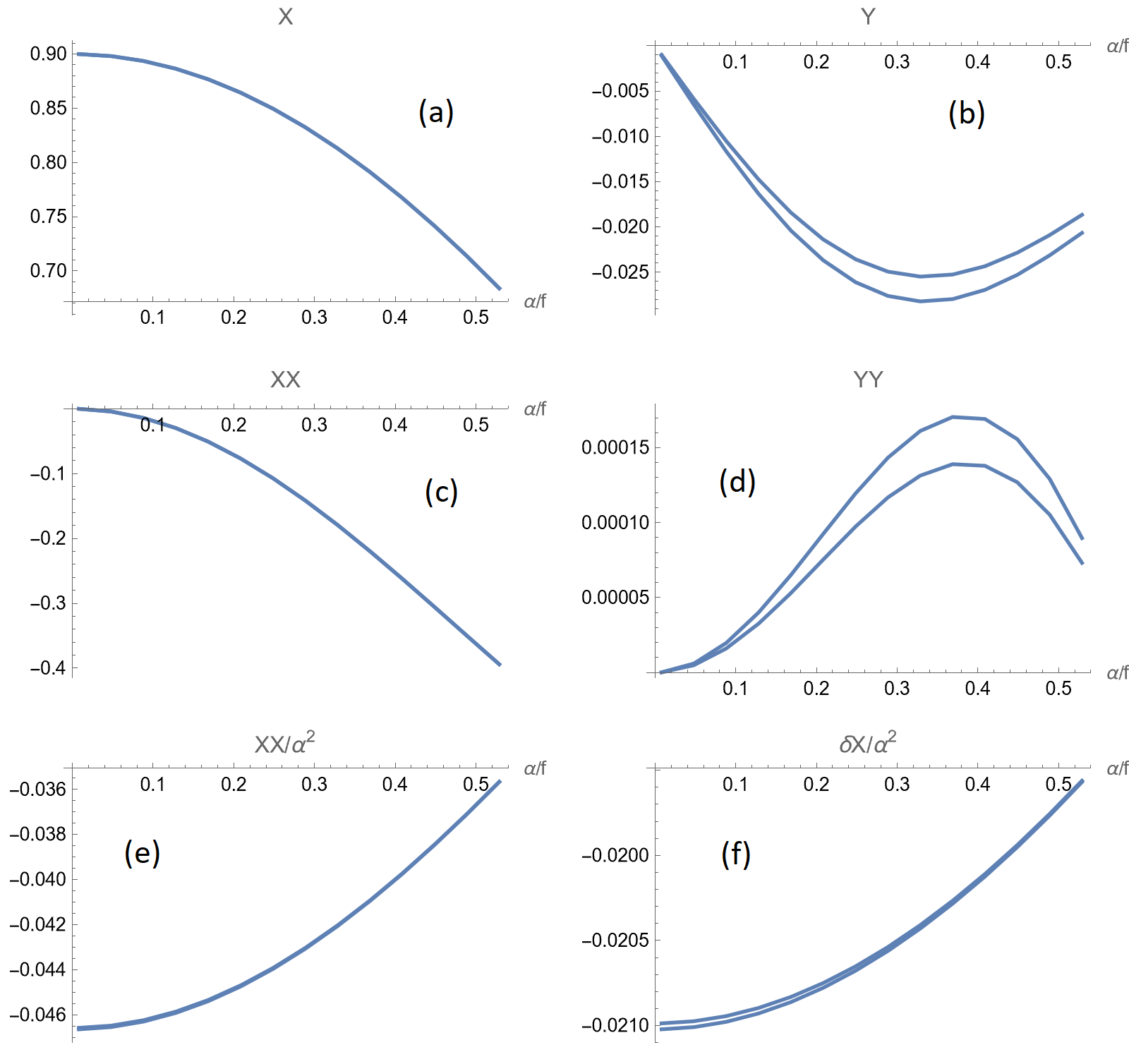}
    \caption{Decomposition of $H_{eff}=-i \log U/P$ into one- and two- nearest neighbor spin terms for $L=8, 10$, with panels (a) and (c) showing terms appearing in leading perturbation theory (Eq. \ref{eq:magnus}) with correct magnitudes, as illustrated in panels (e) and (f). 
    Panels (b) and (d) display terms from beyond the leading order result -- they are rather small, but show the most size dependence. 
    Correction to $ZZ$ term (not shown) is equal and opposite in sign to that of $XX$ term. No $Z$ terms are generated, as expected. The residual missing norm $|H_{eff}^2|-|H_{approx}^2|\lesssim 10^{-4}$ in the entire regime of interest, with weak size dependence.
   \\
   Note: $3/(16 \pi^2)\approx 0.02$. The unitary was computed exactly in Mathematica by multiplying several MatrixExp evaluations (at midpoints of $P/dt=6$ time intervales), with MatrixLog used subsequently to compute $H_{eff}$ -- see text.}
    \label{fig:heff}
\end{figure}
The numerical operation of matrix logarithm, while strictly speaking ill defined (multivalued), may nevertheless be effectively used with care, e.g. when the size of the matrix is small and its spectrum is sufficiently well resolved to preclude close encounters among levels.  In practical terms this means working with small systems, e.g. chains of lengths $L=6..12$, computing and decomposing $H_{eff}$ into sums of local few spin operators. The procedure is justified a posteori if this local approximation to $H_{eff}$ exhausts the numerically exact unitary upon re-exponentiation. All the results presented below were obtained in this fashion for small system sizes and showed good stability while they worked (and they tend to break down for $L\gtrsim 10$ because of multivalued log). Much has been written about the breakdown of Magnus expansion and its possible relationship to thermalization -- the situation we find here (as already anticipated by the analytic calculation), is that $H_{eff}$ remains very nicely local (see Fig. \ref{fig:heff}) even as we venture far from the strict perturbative regime. Most notably, there is a linear in $\alpha$ term that is missing from the third order analytic expansion (it is, in fact, $\sim \alpha J^2 \kappa^2$, i.e. first to show up in fifth order in Magnus expansion!) term that generates a $Y$ transverse field (Fig. \ref{fig:heff}b).  The comparison against perturbative result (Figs. \ref{fig:heff}ef vs. last line in Eq. \ref{eq:magnus}) is imperfect due to the fact that we compute the unitary by subdividing the total period $P$ into steps of duration $dt=P/6$, while analytic results are computed strictly in continuous time. We observed that such small quantitative discrepancies are reduced as $dt$ is decreased.

Note 1:
the convention used in this Supplement differs from the somewhat unusual choice in Main Text (where, following Eq. 3 $|\psi(t+\delta t)\rangle \approx exp[-2\pi i \delta t H(t)]|\psi(t)\rangle $). Instead, we have the more standard $|\psi(t+\delta t)\rangle \approx exp[i \delta t H(t+\delta t/2)]|\psi(t)\rangle$ without the extra $2 \pi$.  As long as $\alpha\to 2\pi \alpha$, the differences in the dynamics appear to be small and insignificant. Importantly, this change turns out to be crucial for the robust implementation of the matrix log, since extra $2\pi$ readily leads to large phases that obscure the log.

Note 2:
Results in this section do not depend on the precise protocol of how $\alpha, f$ are scaled up with $N$.  As explained in main text, scaling both logarithmically appears to be the optimal choice to suppress heating over evolution times relevant to this work. 
For additional theoretical insights into different scaling limits, mostly focusing on steady states, see Ref. \onlinecite{morningstar2023universality}.  In practical terms, such scaling limit should be readily reachable on a digitial quantum computer at the cost of logarithmic increase in circuit depth. Analog implementations of the simulations described in this work may require a more nuanced approach, guided by the general requirement of separation of scales, with heating rates remaining low enough to allow time to observe MQT, which may mean increasing $\alpha, f$ with a modified  protocol guided by physical channels of relaxation specific to the system in use.

\subsection{Dynamics and adiabaticity via Floquet spectra}  
\label{sec:floqadia}
\begin{figure}
    \centering
    \includegraphics[width=\linewidth]{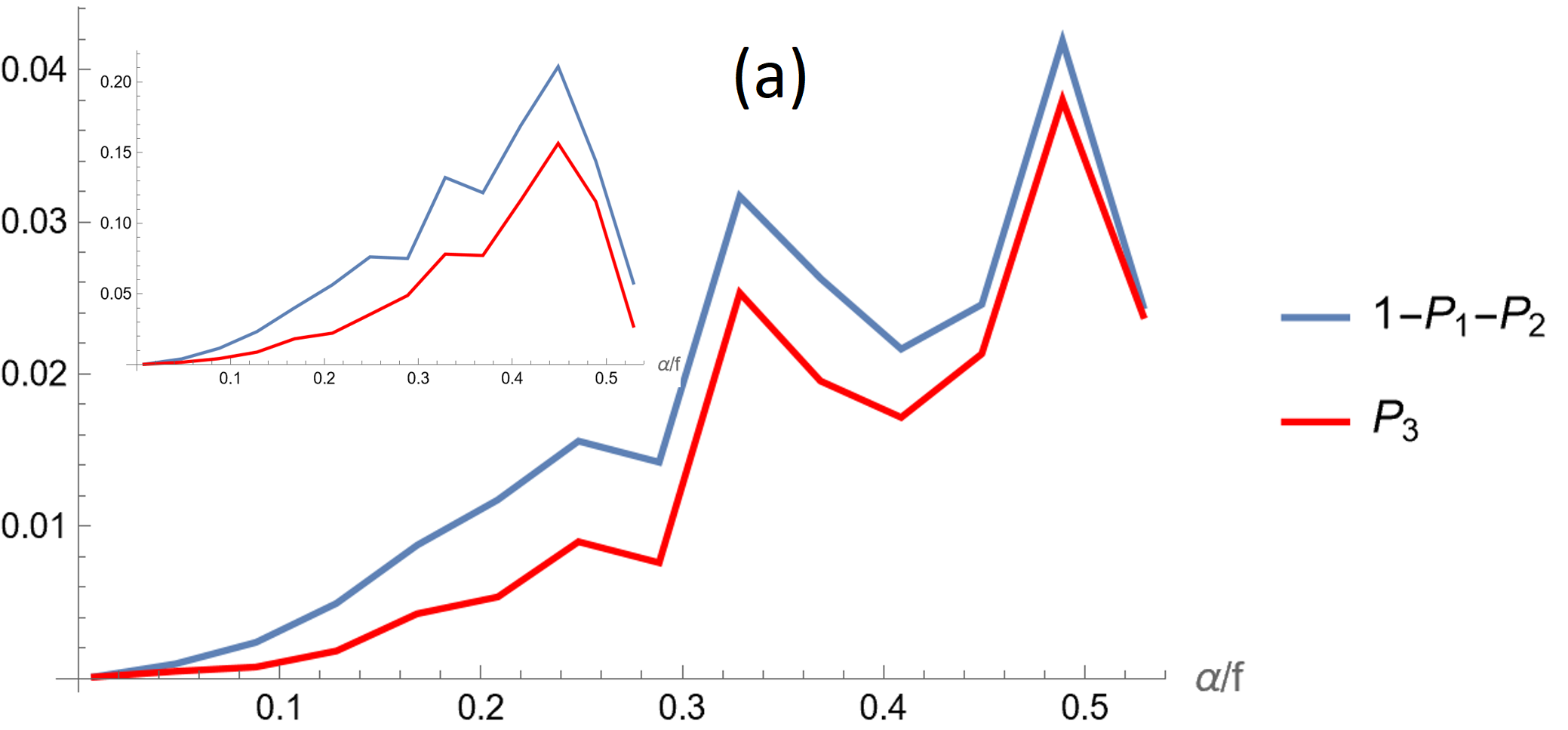}
    \includegraphics[width=\linewidth]{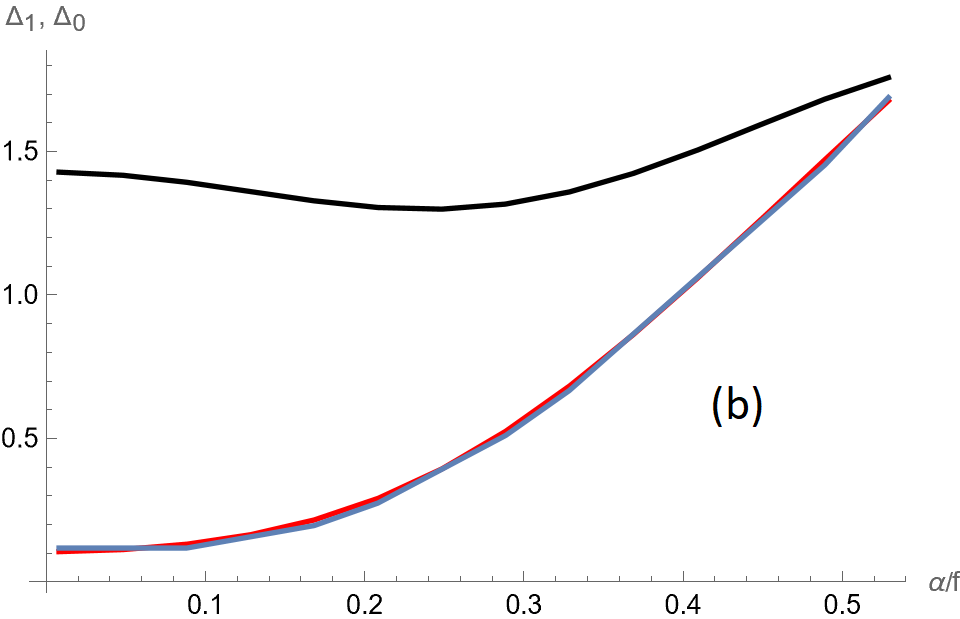}
    \includegraphics[width=\linewidth]{COMBOftZZZ.png}
\caption{(color online) Panel (a): lack of thermalization of Floquet eigenstates with most of the weight saturated by just two eigenstates of the unitary ($P_n$ denotes squared amplitude of state's overlap with a particular eigenstate). Using either of the two quantities plotted as a proxy for diabaticity we note its dependence on dynamic parameters, e.g. $P/dt=6$ (inset) vs.  $P/dt=9$ (main panel);
\\
Panel (b) tracks close agreement between observed frequency of temporal order parameter oscillations (in light blue) obtained from time- and/or Fourier- analysis of order parameter dynamics and the  Floquet gap $\Delta_0$ (in red) between two dominant states (1 and 2, above); Additionally, the gap to the third state $\Delta_1$ is shown in black;
\\
Panels (c) and (d) show Fourier transformed data for average order parameter (in red) and the antipodal correlator (in blue), $\langle Z\rangle(t), \langle Z_0 Z_{L/2}\rangle(t)$, at low frequency and near Floquet zone boundary, respectively, for $\alpha/f\approx 0.2$; Note, while $Z$ has a peak at the collective Rabi frequency $\Omega_R$, the signal in $ZZ$ is at $2\pi/P\pm \Omega_R$;
}
\label{fig:nothermalization}
\end{figure}

As already noted, the effective Hamiltonian we obtained is non-integrable, yet collective oscillations we observe are remarkably coherent and do not show any sign of thermalization.  One natural mechanism for this to happen is if the slowness of the ramp we used is sufficient to deform the initial product state into a linear combination of very few well isolated eigenstates of the Floquet problem (i.e. of $H_{eff}$), that are especially robust, akin to groundstates and low lying excited states of familiar Hamiltonian problems. We now present evidence to support just such a picture, and also flesh out a bit more phenomenology of the fast flipping magnetic state discovered in this work. We only report results for a relatively small TFIM ring of $L=8$ sites, to complement less detailed analysis on larger systems in main text, deferring a proper study of Floquet adiabaticity\cite{schindler2023counterdiabatic} in this problem to future work.

To facilitate the analysis we organize the time by measuring it in units of Floquet period $P=(6 \log 8)^{-1}\approx 0.08$.  Recall, the ramp protocol consists of three stages 
$\{t_r,t_p,t_r\}=2 L \{1,8,1\}\to P\{200,1597,200\}$. Recall, all off-diagonal couplings are turned on smoothly during stage 1 with the envelope function $\sin^2 (0.5 \pi t/t_r)$ (and similarly, turned off in the last stage).  Following analysis above, we will only consider the unitary operator during stage 2 (i.e. only monochromatic drive, no ramps, also no time offset, precisely as in Eq. \ref{eq:Hoft}). In examining the dynamics, however, we will make use of (i.e. peek at!) full  trajectories, including intra-cycle motion. This is referred to as "micromotion" in Floquet literature, although, as we shall see, our problem's micromotion is actually macroscopic! The subdivisions we use are $P/dt=6,9,10$ and usually produce similar results, except as noted in Fig. \ref{fig:nothermalization}a.

\begin{figure}[h]
    \centering
        \includegraphics[width=0.49\linewidth]{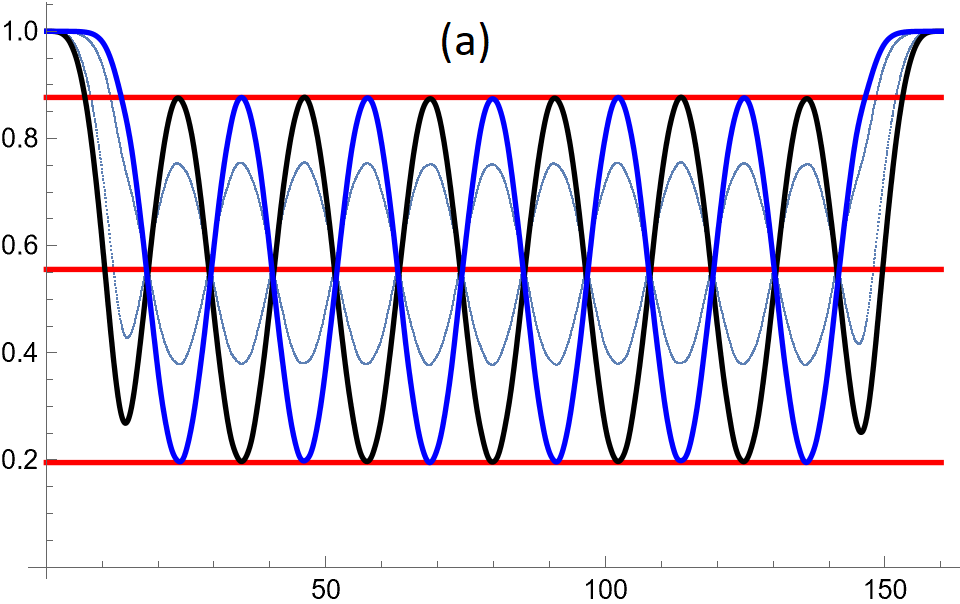}    \includegraphics[width=0.49\linewidth]{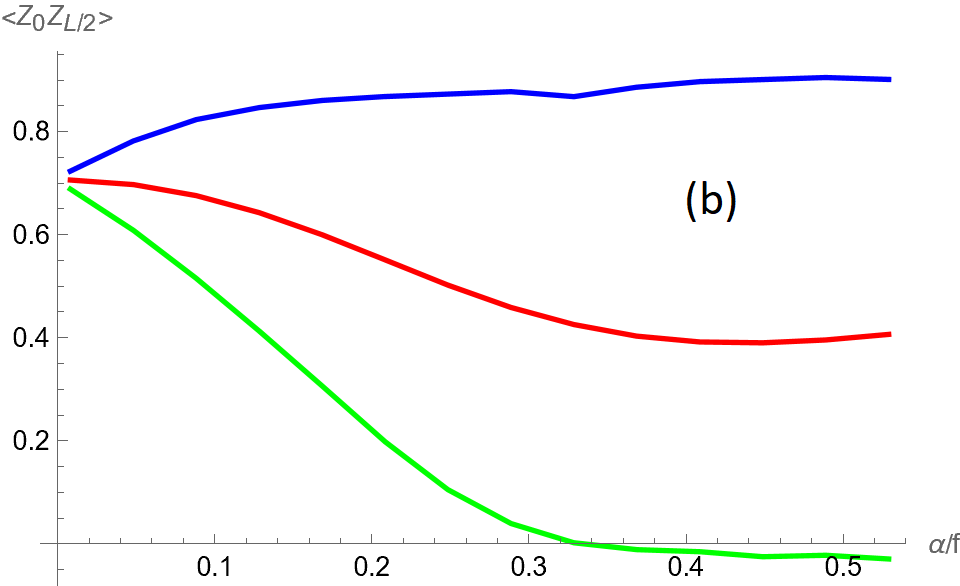}
\includegraphics[width=0.49\linewidth]{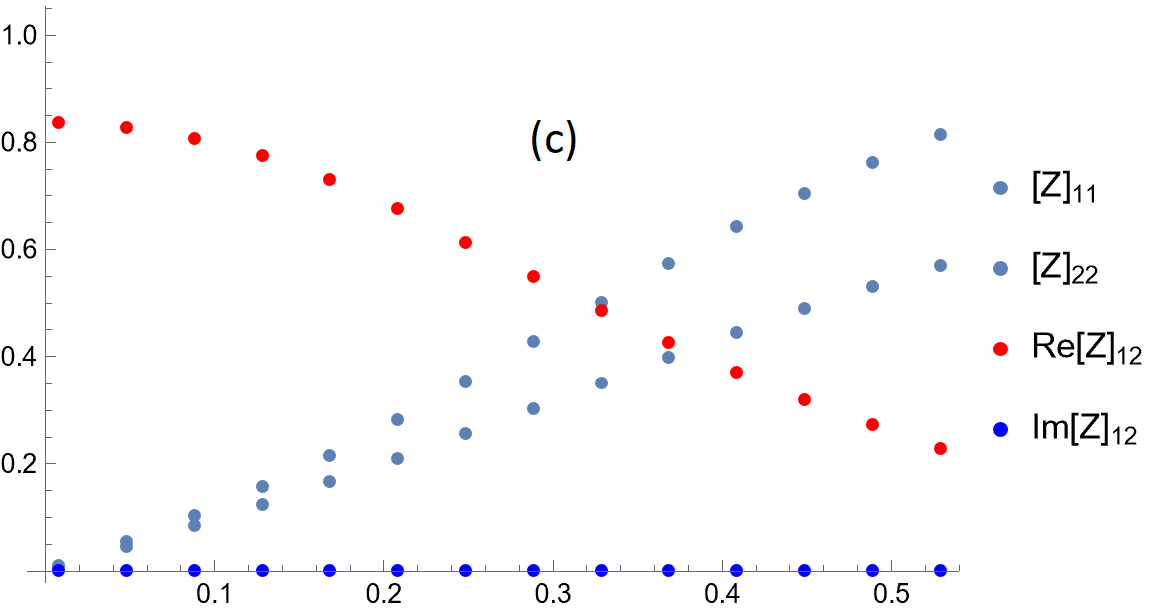}        
\includegraphics[width=0.49\linewidth]{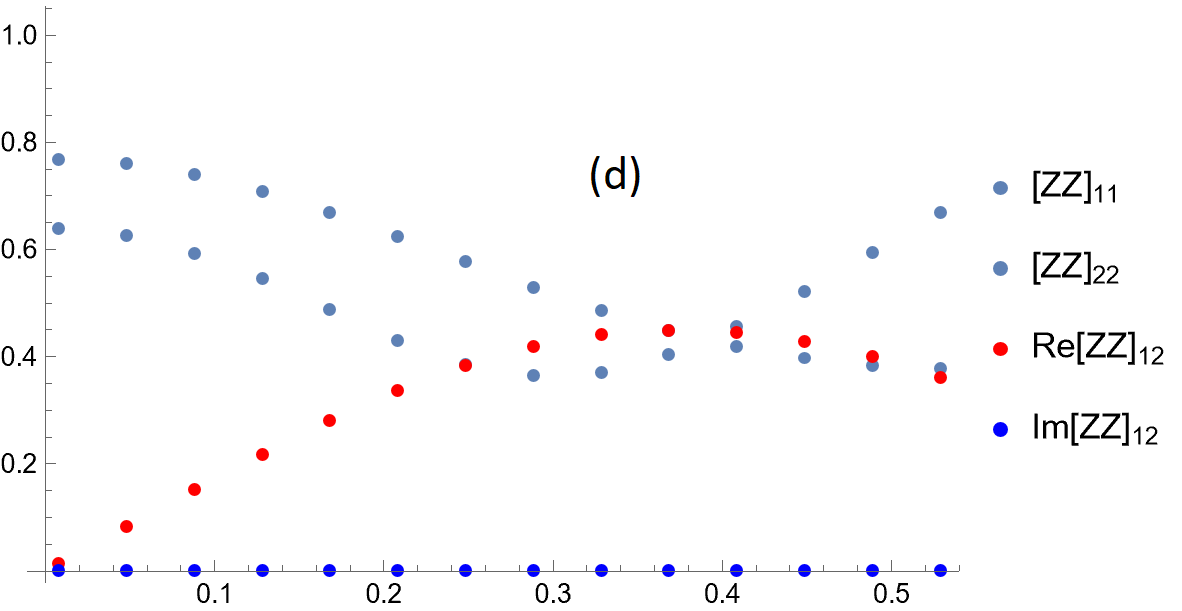}
\includegraphics[width=\linewidth]{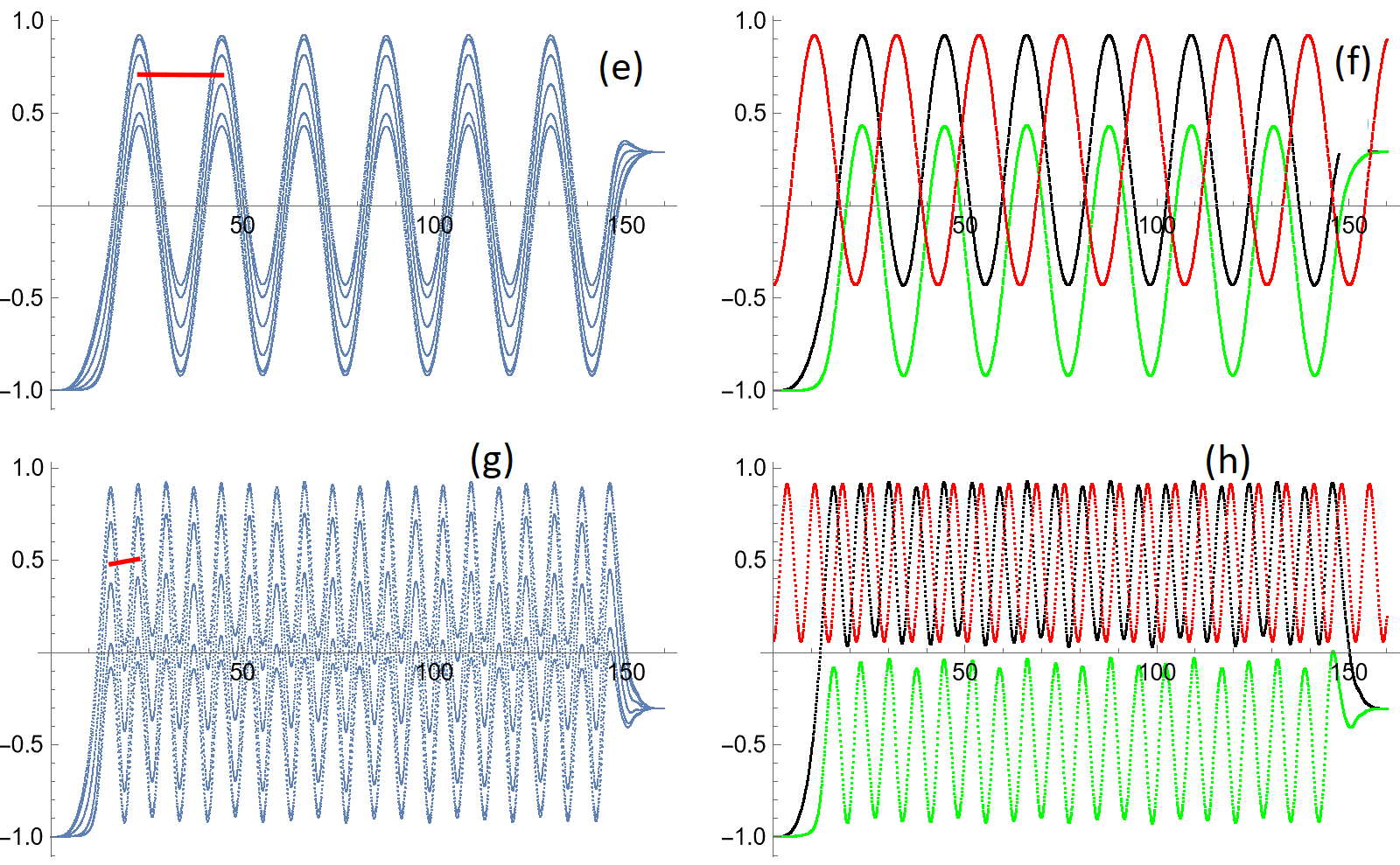}
    \caption{
Real time dynamics and reduced Floquet data: panel (a) shows the dynamics of antipodal correlator at integer period multiples $t=n P$ (black), half-integer multiples $t=(n+1/2)P$ (deep blue) and all other subdivisions $dt$ (light blue), note the nearly perfect anti-correlation; panel (b) shows maximum/average/minimum values of $ZZ$ during stage 2 of dynamcis (in blue/red/green, respectively);
Panels (c) and (d) display the evolution of matrix elements of magnetization and correlator operators within the two dimensional eigenspace isolated in Fig. \ref{fig:nothermalization}a.;
Panels (e) and (g) display full trajectory of mangetization for $\alpha/f\approx 0.2, 0.4$ (respectively); Panels (f) and (h) first repeat the filtering used in panel (a) with integer (black) and half-integer (green) times showing a larger (macroscopic) separation;  We also overlay (in red) an attempt to reconstruct the correlator from spectral data in Figs. \ref{fig:nothermalization}b and panel (c) above -- the match is impeccable (up to an overall phase), unsurprisingly.
}
    \label{fig:realtime}
\end{figure}
We begin by decomposing the time evolved state of the system along the Floquet eigenstates and sorting the resulting probability distribution, to isolate 3 most relevant states, two of which clearly dominate (Fig. \ref{fig:nothermalization}a).  Unsurprisingly, we confirm that the observed oscillations of the order parameter (see Fig.\ref{fig:nothermalization}bc) originate from two-level dynamics between the dominant states. This analysis establishes the high degree of adiabaticity achieved in our protocol. It is noteworthy that the transition to the fast ferromagnetic regime without exponential growth of Rabi frequency discovered and discussed in main text coincides with the approximate softening of the third level near $\alpha/f\approx 0.3$ (Fig. \ref{fig:nothermalization}b), which may be indicative of a quantum critical point and heating effects (some of which we noticed and worked to suppressed in larger $L$ simulations).

We continue the analysis by taking a closer look at the real time dynamics of the order parameter and the correlator in Fig. \ref{fig:realtime}. As a proof of principle we show that observed dynamics is accurately captured in the two level description.
\subsection{Time averaged dynamics}
\begin{figure*}
\includegraphics[width=.66\columnwidth]{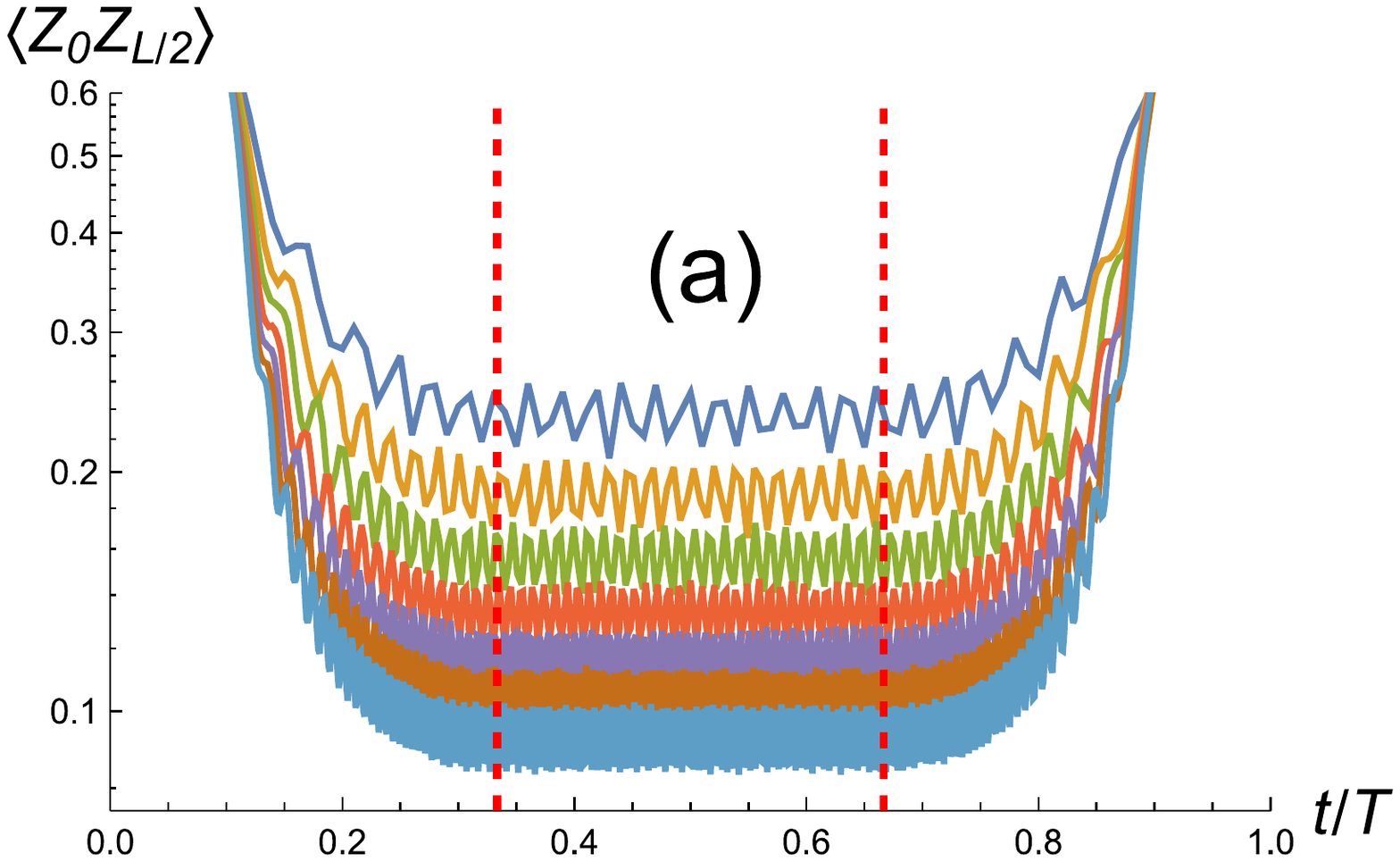}
\includegraphics[width=.66\columnwidth]{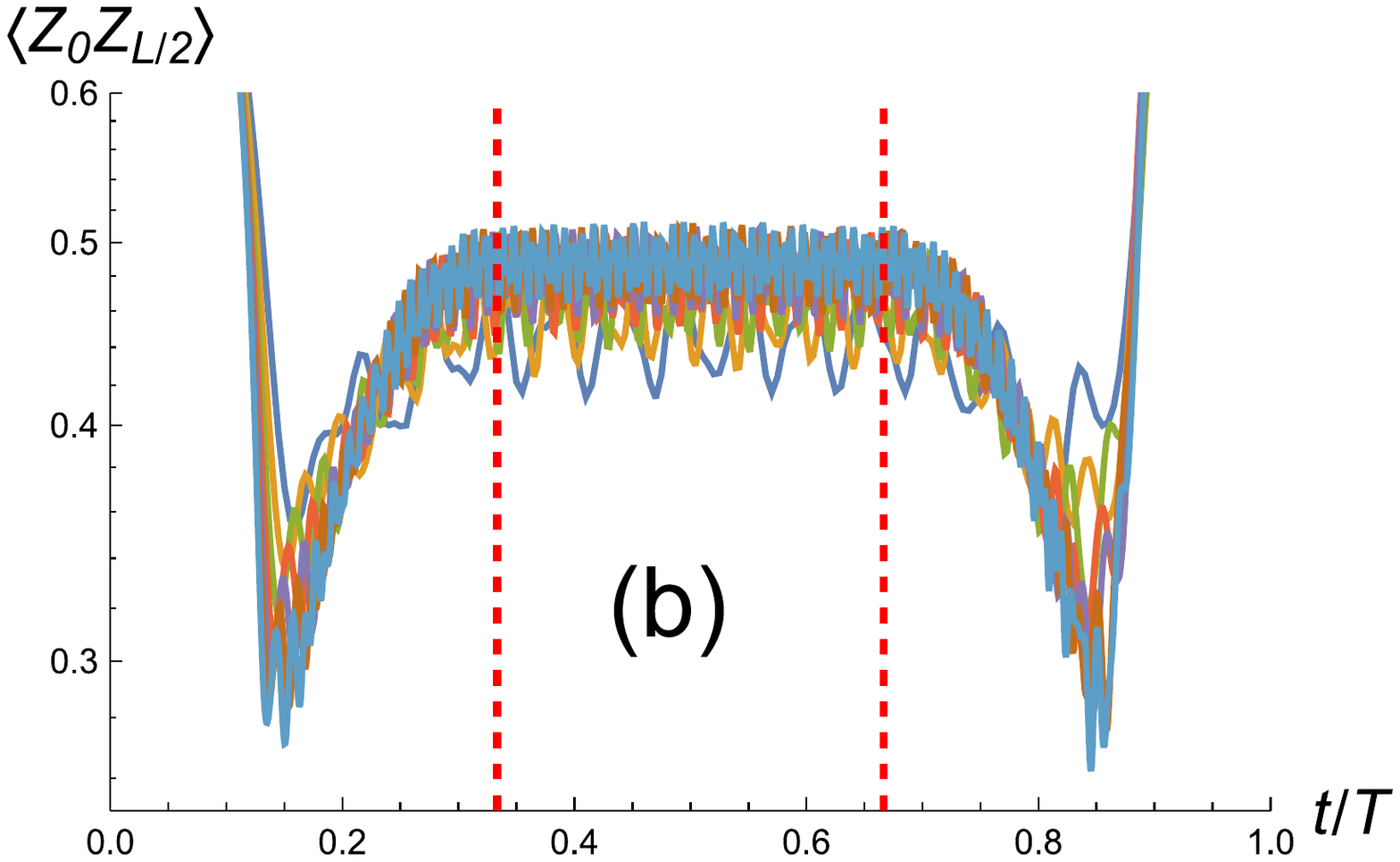}
\includegraphics[width=.66\columnwidth]{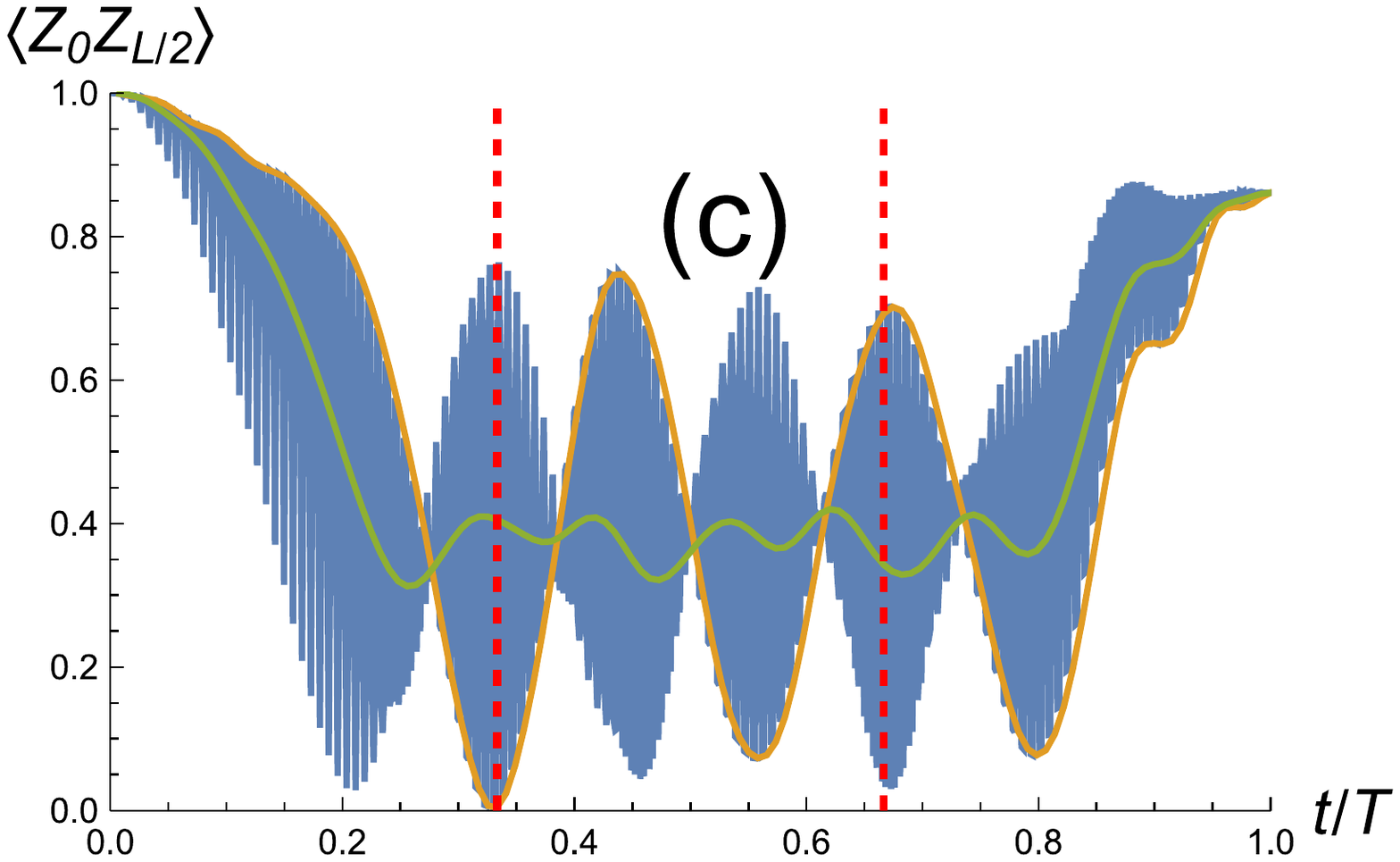}
\caption{Comparison of the time-averaged magnetic order parameter for large DC transverse field (a) and strong AC drives (b), measured over the full evolution time, in the 1d chain with $L$ running from 8 to 20 in steps of 2. Red dashed lines demarcate waiting time in between the ramps. In the DC case on the left, $\kappa$ is ramped up to 2, well past $\kappa_c = 1$, and the residual magnetic order decays exponentially with $L$ as a result. In the AC case on the right, $\kappa$ is ramped up to 0.9 and the AC field is ramped into the strong driving regime with $\alpha/f=7.7/12$. Here the long-range magnetic order is constant with system size in the plateau, but as the fields are ramped up, it does cross a minimum value which is slowly decreasing with $L$, suggesting a phase transition is crossed. $\avg{Z_0 Z_{L/2}}$ begins and ends at 1 (when all transverse fields are turned off) but this region is left off the plot to better focus on the behavior in the ramping and plateau regions.
\\
Panel (c):
sampling dependence of $\avg{Z_0 Z_{L/2}}$ in the strongly driven regime. The blue trace samples it at every timestep, gold at the end of every full period ($dt=P/5$ here), and green sampling at every timestep but reporting the average over each full period. }
\label{fig:DCvsAC}
\end{figure*}

It may be useful and interesting to emulate an imperfect measurement apparatus, that may not be able to resolve the apparently important intra-period dynamics but rather report time averaged results. With this in mind we present results of just such a protocol in Fig. \ref{fig:realtime}, additionally exploring the contrast between the transition from ordinary FM to ordinary PM and the fast flipping FM discovered here. The latter displays a dramatic non-equilibrium recovery effect during the ramp stage 1, which we have not seen in previous figures.

\bibliography{fullbib}